\documentclass[longauth]{aa}  

\usepackage{graphicx}
\usepackage{txfonts}
\usepackage{hyperref}
\hypersetup{colorlinks=true, citecolor=blue, linkcolor=blue}
\begin{document}

    \title{GATOS: missing molecular gas in the outflow of NGC\,5728 revealed by JWST}

\author{
     R.~Davies\inst{1}
\and T.~Shimizu\inst{1}
\and M.~Pereira-Santaella\inst{2}
\and A.~Alonso-Herrero\inst{3}
\and A.~Audibert\inst{4,5}
\and E.~Bellocchi\inst{6,7}
\and P.~Boorman\inst{8}
\and S.~Campbell\inst{9}
\and Y.~Cao\inst{1}
\and F.~Combes\inst{10}
\and D.~Delaney\inst{11}
\and T.~D\'iaz-Santos\inst{12,13}
\and F.~Eisenhauer\inst{1}
\and D.~Esparza~Arredondo\inst{4,5}
\and H.~Feuchtgruber\inst{1}
\and N.M.~F\"orster~Schreiber\inst{1}
\and L.~Fuller\inst{14}
\and P.~Gandhi\inst{15}
\and I.~Garc\'ia-Bernete\inst{16}
\and S.~Garc\'ia-Burillo\inst{17}
\and B.~Garc\'ia-Lorenzo\inst{4,5}
\and R.~Genzel\inst{1}
\and S.~Gillessen\inst{1}
\and O.~Gonz\'alez~Mart\'in\inst{18}
\and H.~Haidar\inst{9}
\and L.~Hermosa~Mu\~noz\inst{3}
\and E.K.S.~Hicks\inst{11}
\and S.~H\"onig\inst{15}
\and M.~Imanishi\inst{19}
\and T.~Izumi\inst{19,20}
\and A.~Labiano\inst{21}
\and M.~Leist\inst{14}
\and N.A.~Levenson\inst{22}
\and E.~Lopez-Rodriguez\inst{23}
\and D.~Lutz\inst{1}
\and T.~Ott\inst{1}
\and C.~Packham\inst{14,19}
\and S.~Rabien\inst{1}
\and C.~Ramos~Almeida\inst{4,5}
\and C.~Ricci\inst{24,25}
\and D.~Rigopoulou\inst{16,13}
\and D.~Rosario\inst{8}
\and D.~Rouan\inst{26}
\and D.J.D.~Santos\inst{1}
\and J.~Shangguan\inst{1}
\and M.~Stalevski\inst{27,28}
\and A.~Sternberg\inst{29,30}
\and E.~Sturm\inst{1}
\and L.~Tacconi\inst{1}
\and M.~Villar~Mart\'in\inst{31}
\and M.~Ward\inst{32}
\and L.~Zhang\inst{14}
    }

\institute{
     Max Planck Institute for Extraterrestrial Physics (MPE), Giessenbachstr.1, 85748 Garching, Germany
\and Instituto de F\'isica Fundamental, CSIC, Calle Serrano 123, 28006 Madrid, Spain
\and Centro de Astrobiolog\'ia (CSIC-INTA), Camino Bajo del Castillo s/n, 28692 Villanueva de la Ca\~nada, Madrid, Spain
\and Instituto de Astrof\'isica de Canarias, Calle V\'ia L\'actea, s/n, E-38205 La Laguna, Tenerife, Spain
\and Departamento de Astrof\'isica, Universidad de La Laguna, E-38205, La Laguna, Tenerife, Spain
\and Departmento de F\'isica de la Tierra y Astrof\'isica, Fac. de CC F\'isicas, Universidad Complutense de Madrid, 28040 Madrid, Spain
\and Instituto de F\'isica de Part\'iculas y del Cosmos IPARCOS, Fac. CC F\'isicas, Universidad Complutense de Madrid, 28040 Madrid, Spain
\and Cahill Center for Astronomy and Astrophysics, California Institute of Technology, Pasadena, CA 91125, USA
\and School of Mathematics, Statistics and Physics, Newcastle University, Newcastle upon Tyne, NE1 7RU, UK
\and LERMA, Observatoire de Paris, Coll\`ege de France, PSL University, CNRS, Sorbonne University,  Paris
\and Department of Physics \& Astronomy, University of Alaska Anchorage, Anchorage, AK 99508-4664, USA
\and Institute of Astrophysics, Foundation for Research and Technology-Hellas, Heraklion, 70013, Greece
\and School of Sciences, European University Cyprus, Diogenes street, Engomi, 1516 Nicosia, Cyprus
\and Department of Physics and Astronomy, The University of Texas at San Antonio, 1 UTSA Circle, San Antonio, TX 78249, USA
\and Department of Physics \& Astronomy, University of Southampton, Hampshire SO17 1BJ, Southampton, UK
\and Department of Physics, University of Oxford, Denys Wilkinson Building, Keble Road, Oxford, OX1 3RH, UK
\and Observatorio Astron\'omico Nacional (OAN-IGN)-Observatorio de Madrid, Alfonso XII, 3, 28014-Madrid, Spain
\and Instituto de Radioastronom\'ia y Astrof\'isica (IRyA), Universidad Nacional Aut\'onoma de M\'exico, Antigua Carretera a P\'atzcuaro \#8701, ExHda. San Jos\'e de la Huerta, Morelia, Michoac\'an, M\'exico C.P. 58089
\and National Astronomical Observatory of Japan, National Institutes of Natural Sciences, 2-21-1 Osawa, Mitaka, Tokyo 181-8588, Japan
\and Department of Astronomy, School of Science, Graduate University for Advanced Studies (SOKENDAI), Mitaka, Tokyo 181-8588, Japan
\and Telespazio UK for the European Space Agency, ESAC, Camino Bajo del Castillo s/n, 28692 Villanueva de la Ca\~nada, Spain
\and Space Telescope Science Institute, 3700 San Martin Drive, Baltimore, MD 21218, USA
\and Kavli Institute for Particle Astrophysics \& Cosmology (KIPAC), Stanford University, Stanford, CA 94305, USA
\and N\'ucleo de Astronom\'ia de la Facultad de Ingenier\'ia, Universidad Diego Portales, Av. Ej\'ercito Libertador 441, Santiago, Chile
\and Kavli Institute for Astronomy and Astrophysics, Peking University, Beijing 100871, People’s Republic of China
\and LESIA, Observatoire de Paris, Universit\'e PSL, CNRS, Sorbonne Universit\'e, Sorbonne Paris Cite\'e, 5 place Jules Janssen, 92195 Meudon, France
\and Astronomical Observatory, Volgina 7, 11060 Belgrade, Serbia
\and Sterrenkundig Observatorium, Universiteit Gent, Krijgslaan 281-S9, Gent B-9000, Belgium
\and School of Physics and Astronomy, Tel Aviv University, Ramat Aviv 69978, Israel
\and Center for Computational Astrophysics, Flatiron Institute, 162 5th Ave, New York, NY 10010, USA
\and Centro de Astrobiolog\'ia (CAB), CSIC-INTA, Ctra. de  Ajalvir, km 4, 28850 Torrej\'{o}n de Ardoz, Madrid, Spain
\and Centre for Extragalactic Astronomy, Department of Physics, Durham University, South Road, Durham, DH1 3LE, UK
    }

    \date{Received ---; accepted ---}
 
\abstract{The ionisation cones of NGC\,5728 have a deficit of molecular gas based on millimetre observations of CO\,(2-1) emission. Although photoionisation from the active nucleus may lead to suppression of this transition, warm molecular gas can still be present. We report the detection of eight mid-infrared rotational H$_2$ lines throughout the central kiloparsec, including the ionisation cones, using integral field spectroscopic observations with JWST/MIRI~MRS.
The H$_2$ line ratios, characteristic of a power-law temperature distribution, indicate that the gas is warmest where it enters the ionisation cone through disk rotation, suggestive of shock excitation. In the nucleus, where the data can be combined with an additional seven ro-vibrational H$_2$ transitions, we find that moderate velocity (30~km~s$^{-1}$) shocks in dense ($10^5$~cm$^{-3}$) gas, irradiated by an external UV field ($G_0 = 10^3$), do provide a good match to the full set. The warm molecular gas in the ionisation cone that is traced by the H$_2$ rotational lines has been heated to temperatures $>200$~K.
Outside of the ionisation cone the molecular gas kinematics are undisturbed.
However, within the ionisation cone, the kinematics are substantially perturbed, indicative of a radial flow, but one that is quantitatively different from the ionised lines. We argue that this outflow is in the plane of the disk, implying a short 50~pc acceleration zone up to speeds of about 400~km~s$^{-1}$ followed by an extended deceleration over $\sim$700~pc where it terminates. The deceleration is due to both the radially increasing galaxy mass, and mass-loading as ambient gas in the disk is swept up.}

    \keywords{Galaxies: active 
            -- Galaxies: individual: NGC\,5728 
            -- Galaxies: kinematics and dynamics 
            -- Galaxies: nuclei 
            -- Galaxies: Seyfert 
            -- Infrared: galaxies}

    \maketitle
%

\section{Introduction}
\label{sec:intro}

\nolinenumbers

It has been known for a number of decades that active galactic nuclei (AGN) drive ionised outflows and that these should play a central role in galaxy evolution \citep{vei05,tom10,fab12}. The importance of molecular outflows has been realised only more recently \citep{stu11,vei13,cic14,fio17,mor17,vei20}.
There is an ongoing effort to assess outflow rates and kinetic powers in order to understand how efficiently these various outflows couple to the interstellar medium and their impact in terms of the host galaxy gas depletion timescale and star formation rate \citep{fio17,rup17,flu19,lut20,flu21,lam22,ram22}.
This is complicated by the inter-dependence of many of the galaxy and AGN properties. These include black hole mass, AGN luminosity, stellar mass, and star formation rate; by the role of geometry, and of the timescales of the processes involved; and by the observational impacts between, for example, type~1 and type~2 AGN, or radiative and mechanical processes.
One outcome of the studies of global properties on scales of kilo-parsecs or more, has been the suggestion that many (perhaps most) molecular outflows remove gas only from the central regions and that this will be re-accreted onto the host galaxy, or equivalently that AGN do not remove gas or quench star formation on a global galaxy scale \citep{ros18,sha18,sha20,flu19,ell21,ram22,lam23,mol23}. 
This points to a need to study these outflows on smaller scales.

Indeed, there has been an extensive parallel effort to attain a more detailed view of the mechanisms at work in the circumnuclear regions, on scales of a few to a few hundred parsecs, to understand how the molecular gas is accelerated.
Such studies are inevitably detailed, and so often focus on individual archetypal objects such as NGC\,1068 \citep{gar14,gal16,imp19,ima20} or outflows associated with a jet as in IC~5063 \citep{mor15,oos17} and the teacup quasar \citep{aud23}. 
In a number of cases, the molecular outflow appears to be radial in the plane of the inner disk \citep{alo18,alo19,alo23,esp24} or the result of an interaction between the ionised outflow and the molecular disk \citep{dav14}.
It is in this context that we study here the molecular gas influenced by the outflow in NGC\,5728, which is a strongly barred disk galaxy at a distance of 39~Mpc.
This galaxy hosts an obscured AGN (optical type~2 and X-ray column $\log{N_H}~(cm^{-2}) = 24.2$) with a bolometric luminosity $\log L_{AGN}$~(erg\,s$^{-1}$)$ = 44.1$ \citep{dav15} and, an Eddington ratio $\gtrsim0.1$ for a black hole mass $\lesssim8\times10^6$~M$_\odot$ \citep{kuo20}.
Based on the measurement of $\sigma*=168$~km~s$^{-1}$ in both the 850~nm Ca triplet feature and 2.3~$\mu$m CO~(2-0) bandhead \citep{cag20}, this puts the AGN well below the M$_{BH}-\sigma*$ relation.

The disk of NGC\,5728 is generally well detected in the CO\,(2-1)~230.5\,GHz line out to radii of 1.5--2~kpc \citep{shi19,shin19}.
It is remarkable, however, that CO emission is not detected in the ionisation cone.
\cite{shi19} suggested that here the AGN photoionisation may lead to suppressed CO\,(2-1) line emission.
 This phenomenon has been reported for other objects such as NGC\,2110 \citep{ros19} and ESO~428-G014 \citep{fer20}, in both of which the molecular gas is revealed by the presence of near-infrared H$_2$ lines tracing the hot component of molecular gas.
At smaller scales, different causes for weak CO\,(3-2) or CO\,(2-1) emission may be at play. 
In the central 100~pc of NGC\,3227, HCN\,(1-0) was associated with faint CO\,(2-1) and explained in terms of high HCN abundance due to X-ray irradiation of the gas by the AGN \citep{dav12}.
For several objects, in the central few tens of parsecs where there is only faint CO\,(3-2) emission, the presence of significant molecular gas has been revealed by HCO$^+$\,(4-3) and explained in terms of density stratification \citep{gar21}. 
And in Circinus the weak CO\,(3-2) in the nucleus has been explained in terms of self-absorption, and the presence of molecular gas is revealed via the higher CO\,(6-5) transition as well as other higher density tracers \citep{tri22}.

For the outflow in NGC\,5728 on scales of 100--1000~pc, near-infrared H$_2$ lines have been detected in the outflow, suggesting that molecular gas may be present \citep{dur18,shi19}.
However, these ro-vibrational lines trace gas at typically 1000--2000~K, which comprises a very small fraction of the total mass.
It is therefore expedient to look at the pure rotational H$_2$ lines present at mid-infrared wavelengths, that are produced by warm molecular gas at temperatures of 100--1000~K and which are expected to trace a larger fraction of the total molecular mass.
We present an analysis of such observations in this paper.

\begin{figure*}
    \centering
    \includegraphics[width=15cm]{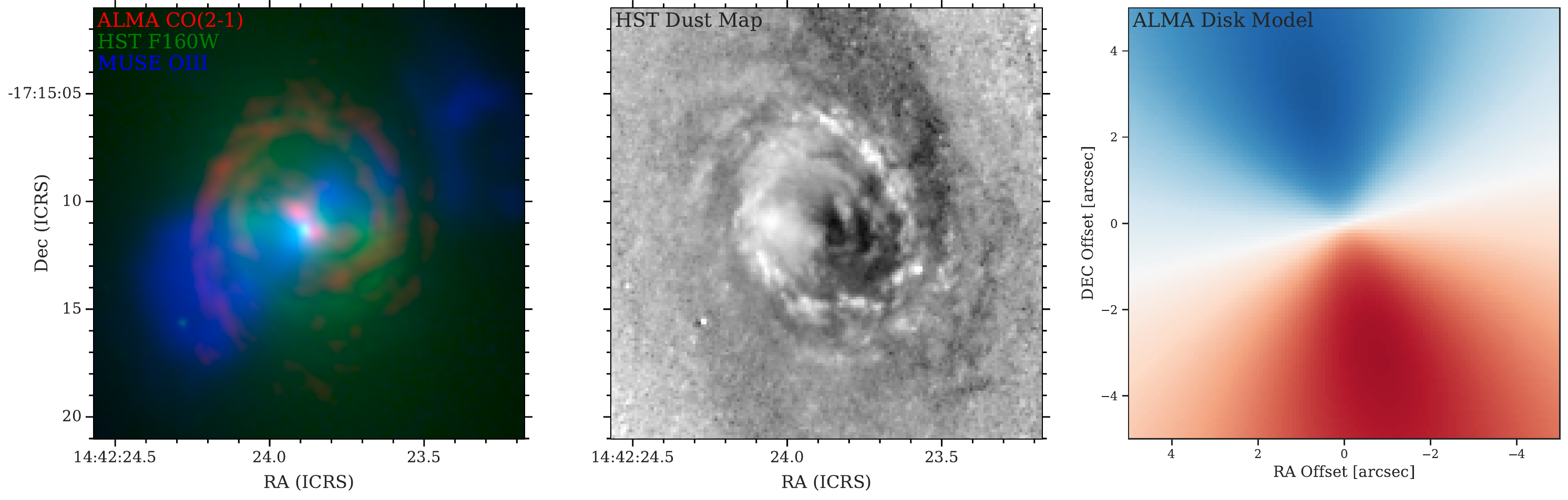}
    \includegraphics[width=15cm]{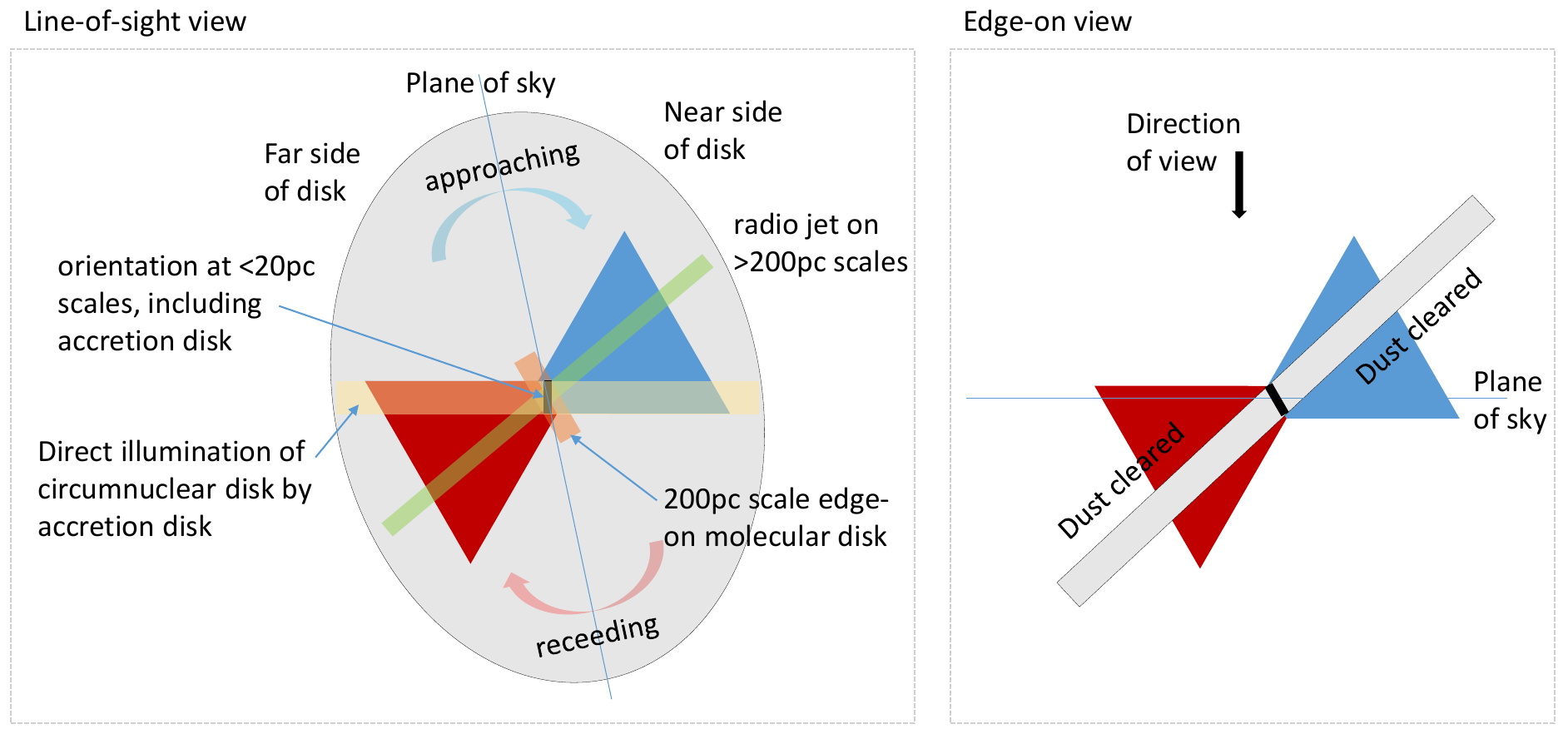}
    \caption{The circumnuclear region of NGC\,5728. 
    Top left: three colour montage of the ALMA CO\,(2-1) map (red), HST F160W H-band continuum (green), and VLT/MUSE [OIII] line emission (blue). These show several key components: the ionisation cone, and the molecular gas that is collimating it, and the direct illumination of the circumnuclear disk by the AGN. 
    Top centre: B-H dust structure map (from F438W and F160W HST images; there is no strong line contamination in these filters) showing the dusty circumnuclear disk to the west and a lighter region to the east where the outflow has removed dust from our line of sight.
    Top right: Velocity field of the disk model fitted to the ALMA CO\,(2-1) kinematics.
    Bottom: cartoon showing two views to illustrate the complex geometry, highlighting that the ionisation cone strongly intersects the galaxy disk (similar to scenario~A in \citealt{ram22}).
    All panels are adapted from, or based on the kinematic modelling of, \citet{shi19}. The cartoon has been slightly updated to better match the spatial distributions.}
    \label{fig:circumnuclear}
\end{figure*}

\section{The circumnuclear region of NGC\,5728}
\label{sec:circum}

As a context to the study presented here, we summarise the relevant points concerning what is known about the circumnuclear region of NGC\,5728. Because it can be well resolved spatially at 100~pc scales or less with 8-m class telescopes at optical and near-infrared wavelengths, it has been studied extensively \citep{son09,shi19,dur18,dur19,shin19}.
A focus of all these studies has been the prominent star forming ring at a radius of about 800~pc and the ionisation cone, which has also been detected in soft X-rays extending to beyond 1~kpc \citep{tri23}.
These primary features are seen in the top left panel of Fig.~\ref{fig:circumnuclear} and mean that the central kiloparsec of this galaxy is a complex region both morphologically and kinematically.

The top right panel shows a $B-H$ dust structure map, which highlights several key features.
Inside the star forming ring, the bright and dark arcs on the western side trace the dusty circumnuclear disk.
But the eastern side is very different, dominated by a conical region that is pale, indicating that the dust has been driven out (indeed, this is a typical signature of the interaction of an outflow with the surrounding disk, see for example \citealt{dav14}).
While there have been previous suggestions that there may be a circumnuclear stellar bar, based on HST imaging, \citet{wil93} noted that the features could also be scattered nuclear light, which would be more consistent with the dust structure map and also the lack of evidence for a corresponding gaseous bar \citep{pet02}.
We propose a cause for the latter interpretation in Sec.~\ref{subsec:astrom}, and argue below that it is also more consistent with the kinematics.

The inflow driven by the large scale bar terminates at its inner Lindblad resonance, at which radius the gas piles up and inevitably forms stars. 
The kinematics of the disk inside this region are driven by the rotating potential of the bar, which can stimulate a circumnuclear spiral \citep{mac04a,mac04b}.
As pointed out by \citet{can93}, and shown by \citet{dav09} in the context of circumnuclear spirals, an $m$-arm density perturbation appears as an ($m-1$)-arm kinematic wave in the projected line-of-sight velocity field.
Thus, when one removes the circular component of the velocity field (e.g. by subtracting a disk model), the projection of the inflow/outflow associated with a 2-arm spiral would appear as a 1-arm kinematic spiral.
Indeed, the structure in the residual velocity field  for both the molecular gas and the stars does resemble a low amplitude 1-arm spiral (Fig.~9 of \citealt{shi19}, and also the bottom right panel of Fig.~\ref{fig:h2_maps} here).
We therefore conclude that the kinematics of the disk on scales of about 0.1-1~kpc are consistent with a circumnuclear spiral stimulated by the large scale bar.

There is another structure in the inner 200~pc that is reminiscent of a nuclear bar: a feature in the CO\,(2-1) map perpendicular to the outflow direction (Figs.~3 and~12 in \citealt{shi19}). However, as argued through the decomposition performed in that work and shown through higher resolution data (Garc\'ia-Burillo et al., in prep), this is in fact a more highly inclined disk on small scales that delineates the plane from which the outflow is emerging.
Geometric modelling of the kinematics by \citet{shi19} indicate that the approaching cone is oriented only 13\degr\ behind the near side of the disk, while the receeding cone is 13\degr\ in front of the far side of the disk. The modelled opening angle of $\pm23$\degr\ means that the cone intersects the disk substantially. This is illustrated in Fig.~\ref{fig:circumnuclear} where we have modified the parameters slightly to better match the spatial distributions, rather than kinematics alone, so that the cone is oriented at a position angle of $-60$\degr\ with an opening angle of $\pm30$\degr.
The inner 200~pc scale disk, because it is closer to edge-on, is likely to be a source of the optical obscuration and X-ray absorbing column that make NGC\,5728 a Compton thick Seyfert~2.

The green line in Fig.~\ref{fig:circumnuclear} denotes the direction of the jet that has been reported at both 6~cm and 20~cm, which shows a spectral index typical of synchrotron rather than thermal emission, and which is thought to be interacting with the disk \citep{sch88,dur18}.
The 20~cm image from the NRAO VLA Archive Survey, for which the beam is $2.0\arcsec\times1.2\arcsec$ at a PA of -2\degr, highlights that on scales of at least a few hundred parsecs the jet is well aligned with the ionisation cone, and is 1-sided (a reverse jet would have to be at least a factor 10 lower surface brightness at this resolution).

We end this short summary of the circumnuclear region of NGC\,5728 with a look at the molecular gas concentration index, or deficit, put forward by \cite{gar21}. 
They compared the ratio of molecular mass surface densities within 50~pc ($\Sigma_{H_2}^{50pc}$) to that at 200~pc ($\Sigma_{H_2}^{200pc}$), to the X-ray luminosities.
The index $\log{\Sigma_{H_2}^{50pc}/\Sigma_{H_2}^{200pc}} = 0.04$ (Garc\'ia-Burillo et al. in prep) and luminosity of NGC\,5728 put it clearly within the locus of AGN which are in the process of removing molecular gas that has built up in their central regions, and so have a flat rather than cuspy radial surface density distribution.
This is consistent with its location in the comparison of Eddington ratio to molecular gas column density in \citet{alo21}. Matching the CO-to-H$_2$ conversion factor used in their analysis, and with the lower limit on Eddington ratio noted above, one finds NGC\,5728 in the locus where dusty outflows are expected to be launched.
It is important to bear in mind that these comparisons are independent of the detailed morphology, and in particular that the concentration index is comparing gas content on very small scales -- within radii of 0.5\arcsec\ and 2\arcsec\ for NGC\,5728.
Thus, the lack of CO\,(2-1) in the ionisation cones, which is the rationale for the analysis here, is part of the variance between galaxies that the index is designed to minimize; and what the index is tracking is the removal of gas from the inner 200~pc scale edge-on disk introduced above.

\section{Sample and target selection}
\label{sec:sample}

NGC\,5728 is one of six galaxies observed in JWST Cycle~1 programme \#1670. 
They form part of the larger sample of AGN in the Galactic Activity, Torus, and Outflow Survey (\href{https://gatos.myportfolio.com}{GATOS}) which are drawn from the 70~Month {\it Swift}-BAT All-sky Hard X-ray Survey \citep{bau13}.
This ensures a nearly complete selection of AGN with 
luminosities $L_{14-150keV} > 10^{42}$\,erg\,s$^{-1}$ at distances of 10--40\,Mpc.
The selection band also means that the sample is largely unbiased to obscuration/absorption even up to column densities of $N_H \sim 10^{24}$\,cm$^{-2}$.
Both the AGN luminosity and absorbing column are available from the analysis of the X-ray data \citep{ric17}.
The GATOS sample, and analysis of the first sub-samples, are described in \citet{gar21} and \citet{alo21}; and a first analysis of the JWST data, focussing on silicate features and water ice at small scales, is presented by \citet{gber24}.
Additional papers about this sample include 
Zhang et al. (in prep) giving on overview of the sample, 
Hermosa~Mu\~noz et al. (in prep) on NGC\,7172, 
Esparza~Arredondo et al. (in prep) on MCG-05-23-016, 
Haidar et al. (in prep) on ESO\,137-G034, 
Garc\'ia-Bernete et al. (in prep) on the spatially resolved PAHs, and 
Lopez-Rodriguez et al. (in prep) on line contamination of continuum images and the dust distribution.

The sub-sample from which NGC\,5728 is selected were also part of the Luminous Local AGN and Matched Analogues (LLAMA) survey \citep{dav15} which were drawn from the same parent sample. 
Observations of these galaxies include integral field spectroscopy at optical and near-infrared wavelengths, the latter using adaptive optics to achieve higher spatial resolution.
In particular, they are included in an analysis of the density of ionised outflows \citep{dav20} and so ionised outflow rates calculated in a consistent way are available for all of them.
This analysis was only performed for Seyfert types 1.8-2, and so also limits the impact of differing orientations.
The sub-sample spans a limited range in distance, stellar mass, and AGN luminosity; but a wide range in ionised outflow rate measured on 150\,pc scales.
As part of the LLAMA survey, NGC\,5728 was the focus of a combined analysis of CO\,(2-1) data from the Atacama Large Millimeter/submillimeter Array (ALMA) wide field optical integral field spectroscopy from the Very Large telescope (VLT) MUSE and high spatial resolution integral field spectroscopy from VLT/SINFONI \citep{shi19}. These data are included in the analysis here to supplement the JWST observations.

\section{Observations, data processing, map extraction}
\label{sec:obs}

Observations of NGC\,5728 were performed on 13--14~March~2023 using the Medium Resolution Spectrometer (MRS; \citealt{wel15,arg23}) of the Mid-infrared Instrument (MIRI; \citealt{rie15}) on the JWST \citep{gar23}.
This instrument offers integral field spectroscopy between 4.9 and 27.9\,$\mu$m at a resolution of R$\sim$1500 to $\sim$3500, over a field of view from $3.2\arcsec\times3.9\arcsec$ at the shortest wavelengths to $6.6\arcsec\times7.7\arcsec$ at the longest.
The full wavelength range comprises 12 segments divided across four channels. To observe the complete range required three sets of observations.

The observations consisted of a $2\times2$ dithered mosaic covering a minimum common field of about $6.6\arcsec\times7.9\arcsec$ oriented along the direction of the ionised outflow, and covering part of the circumnuclear star-forming ring. There was a single integration of 263\,s at each of the four dither positions and for each of the short, medium, and long wavelength settings.
The target observations were accompanied by a pair of dithered background exposures, offset by about 2\arcmin.

\begin{figure}
    \centering
    \includegraphics[width=8cm]{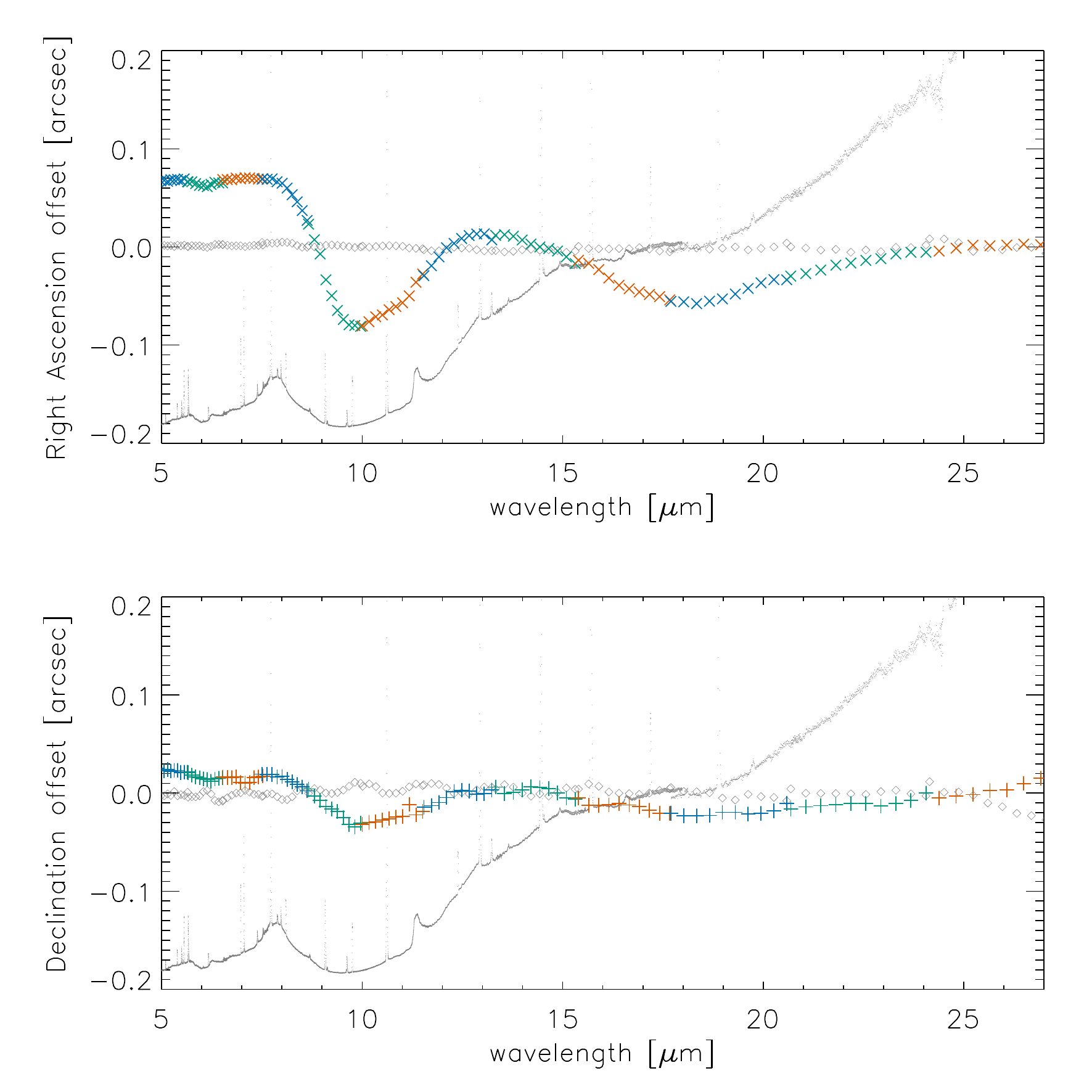}
    \caption{Relative position of the continuum peak of NGC\,5728 in right ascension (top) and declination (bottom). The colours demarcate the spectral channels, showing that the datacubes are well aligned spatially and that there is a wavelength dependent shift of the position. For reference, the equivalent measurement for MCG-05-16-023, dominated by a continuum point source and, as is more typical of the sample, showing no spectro-astrometric shifts is also shown as open diamonds (see Esparza-Arredondo et al. in prep); and the nuclear spectrum is overplotted.}
    \label{fig:astrom_offset}
\end{figure}

\subsection{Data reduction}
\label{subsec:data}

The MIRI MRS data were processed using the JWST Calibration pipeline (version 1.11.4) with the
calibration context 1130. We followed the standard MRS pipeline procedure adding some extra steps to correct for bad pixels.  We estimated the background emission by calculating the median background level at each wavelength channel in the dedicated background data cubes. This median value for each spectral channel was subtracted from the science data cubes. The data reduction is described in detail in \citet{gber22}, \citet{gber24}, and  \citet{mps22}.

In the pipeline data cubes, we found a relatively large offset of 0.27\arcsec\ in right ascension and 0.06\arcsec\ in declination between the mid-infrared continuum peak and the 1.3~mm continuum position measured with a 0.5\arcsec\ beam using ALMA \citep{shi19}. 
Therefore, we corrected the astrometry of the 12 individual MIRI MRS datacubes using background galaxies detected in the parallel imaging. After applying this correction, we found that the relative centering precision was about 0.02\arcsec\ {\sc rms}, which could be verified by tracing the peak position of the AGN as a function of wavelength across all the datacubes. The positions at adjacent wavelengths in differing cubes were matched to the same precision.  

\subsection{Spectro-astrometric shifts}
\label{subsec:astrom}

Interestingly, as shown in Fig.~\ref{fig:astrom_offset}, there were significant systematic excursions of the peak position through the cubes, which is in stark contrast to the equivalent measurement in other objects observed with the same set-up such as MGC-05-23-016 which is dominated by a point-like continuum source (\citealt{asm14}; Esparza-Arredondo et al. in prep.) and the peak position of which shows no measurable shift over the 5--28~$\mu$m wavelength range.
In NGC\,5728 there is a shift of 0.1\arcsec\ close to the east-west direction centered on the 10~$\mu$m silicate absorption feature, and there are secondary shifts of 0.05\arcsec\ at $<8$~$\mu$m and around 18~$\mu$m. 
These indicate that the central peak has spatial structure on scales of 0.1\arcsec, equivalent to $\sim$20~pc.
The position shifts may be indicative of significant obscuration on those scales as similarly proposed by \cite{dia11} for ultraluminous infrared galaxies.
The scale and direction indicate that it is associated with a sub-structure inside the 200~pc edge-on disk discussed in Sec.~\ref{sec:circum}, perhaps related to the third component identified by \cite{shi19}. It points to another twist of the disk on smaller scales. A similar situation, causing a misalignment between the accretion disk and the outflow direction, has been demonstrated for Circinus \citep{sta17,sta19,sta23}.
In the case of Circinus, it leads to an asymmetric illumination of the dust at the edges of the ionisation cone.
Here, the direction of the obscuration gradient implies there is a structure such as a disk with its projected major axis aligned north-south, and tilted a little away from edge-on so that the western side points slightly towards us and is less obscured, while the eastern side is more obscured.
This is illustrated as the innermost black line in the cartoon of Fig.~\ref{fig:circumnuclear}.
Such a geometry would provide a natural explanation for the bar-like brightening in the continuum images: this is instead direct illumination by the accretion disk of the circumnuclear disk at one edge of the ionisation cone.
Similar intrinsic anisotropic illumination of the circumnuclear region due to a small scale warped disc was proposed by \citet{tad00} for Cygnus\,A.
We leave it as a proposed geometry because the detailed modelling needed to explore the scenario more quantitatively, including how it creates this astrometric signature, and how the sub-pc geometry from the interpretation of the maser measurements as a magnetocentrifugal wind \citep{kuo20} fits in, is beyond the scope of this work.

\begin{figure*}
    \centering
    \includegraphics[width=13.5cm]{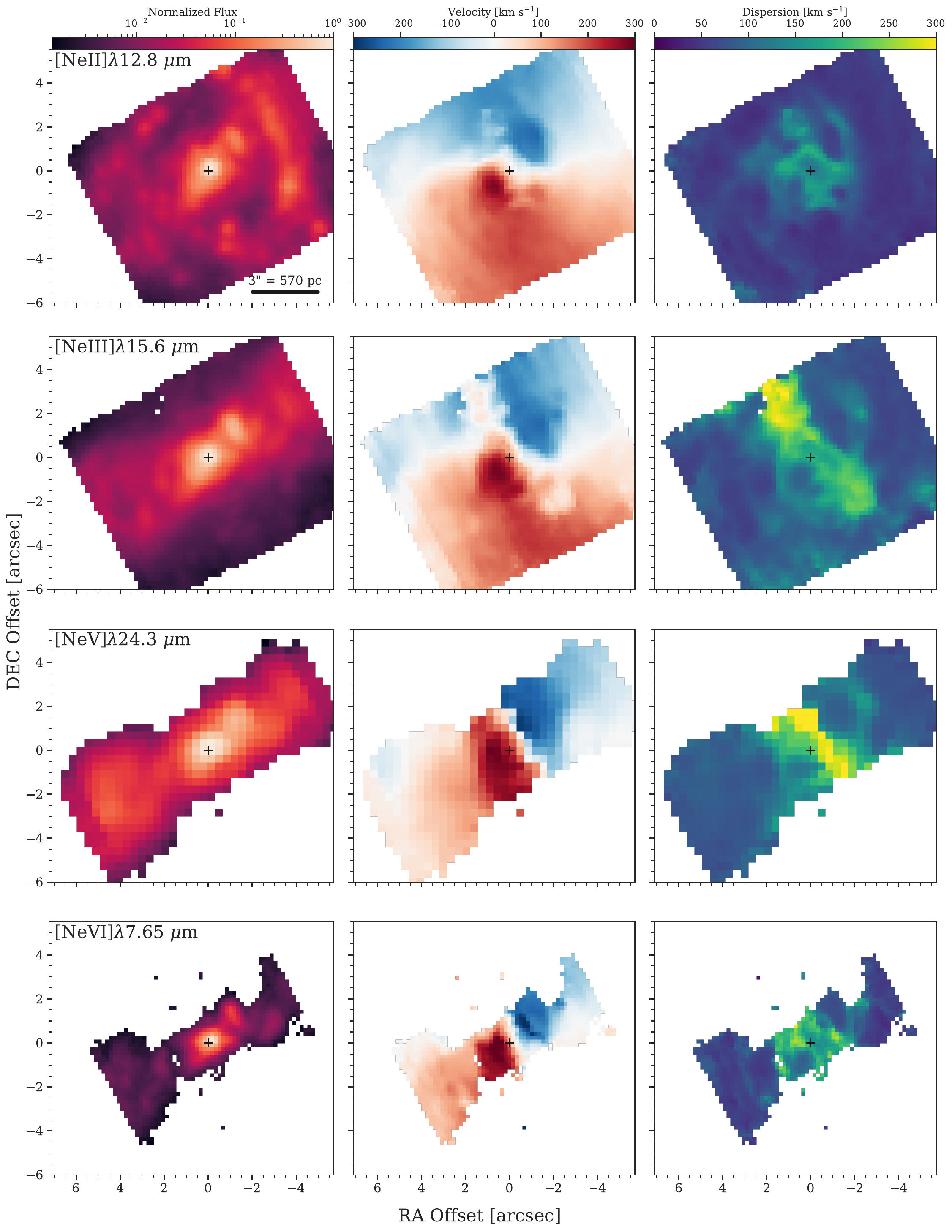}
    \caption{Flux and Kinematics maps of four neon lines (north is up, east is left). The flux (left), velocity (centre), and velocity dispersion (right) of the [Ne\,II]~12.8~$\mu$m, [Ne\,III]~15.6~$\mu$m, [Ne\,V]~24.3~$\mu$m, and [Ne\,VI]~7.7~$\mu$m transitions are shown to provide a context about the circumnuclear region of NGC\,5728. As the ionisation potential increases from 22~eV to 126~eV (top to bottom), what the line emission traces changes from star formation in the ring to the outflow. The direction of the major kinematic axis and the central dispersion change correspondingly. Interestingly, the outflow kinematics reveal anomalous velocities and unusually high dispersion along the minor axis, which is especially noticeable in the [Ne\,III] line.}
    \label{fig:ne_maps}
\end{figure*}

\begin{figure*}
    \centering
    \includegraphics[width=18cm]{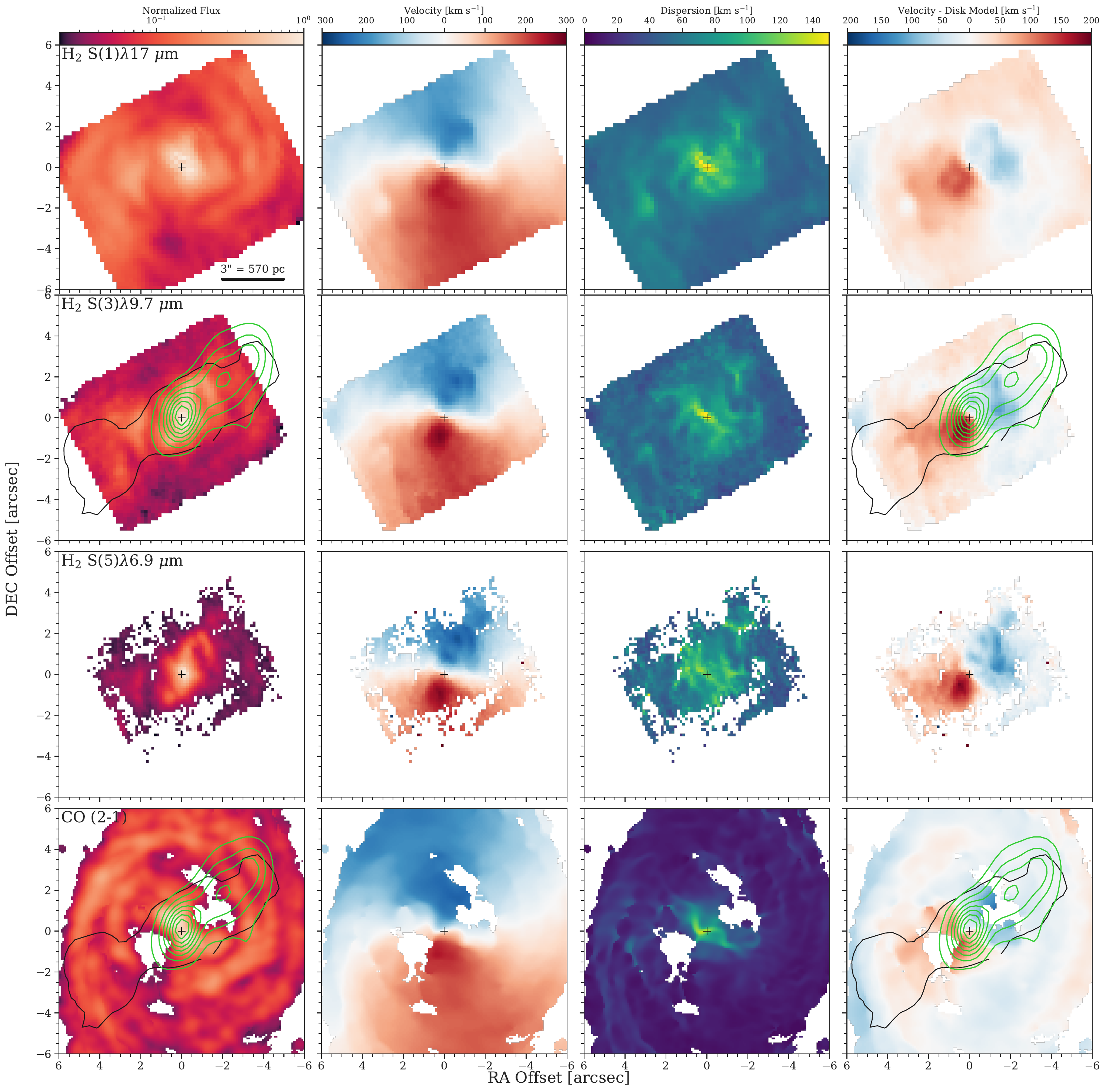}
    \caption{Flux and Kinematics maps of molecular lines (north is up, east is left). The flux (far left), velocity (centre left), velocity dispersion (centre right), and residual velocity after subtracting a disk model (far right) are shown for the H$_2$ 0-0\,S(1), 0-0\,S(3), and 0-0\,S(5) and 1.3\,mm CO\,(2-1) lines. The CO data and disk model are from \cite{shi19}. These data are the focus of the analysis in Sec.~\ref{sec:h2temp} and Sec.~\ref{sec:h2_kin}. The most striking features are: (i) the morphology and kinematics of the CO and S(1) lines are remarkably similar indicating they trace the same gas component, (ii) while CO is not detected in the ionisation cones, the rotational H$_2$ lines are detected with a perturbed velocity field, and (iii) the largest velocity residuals are well-aligned with the ionisation cone as traced by the [Ne\,V] emission (black contour) and the radio jet at 20~cm (green contours).}
    \label{fig:h2_maps}
\end{figure*}

\subsection{Flux and kinematics maps}
\label{subsec:maps}

Flux and kinematics maps for each line were extracted using a single Gaussian fit to the line profile and a linear continuum to match the local continuum.
A quadrature correction was applied to the velocity dispersion to account for the instrumental spectral resolution. 
In order to provide context about the circumnuclear region, the resulting maps for four neon lines are shown in Fig.~\ref{fig:ne_maps}. While these are not a central part of the analysis here, and so are not discussed in detail, we describe the main features that they reveal.

The [Ne\,II]~12.8~$\mu$m line with an ionisation potential\footnote{The term `ionisation potential' can be ambiguous when applied to emission lines. In some cases an emission line arises following recombination from the next higher ionisation level. Here we are considering fine structure lines, which originate from collisional excitation with electrons, and so the ionisation potential refers to the ion's current state.} (IP) of 21.6~eV traces star formation in the circumnuclear ring, where the kinematics are dominated by the disk rotation \citep{shi19}. 
In the central region, the twist in the zero velocity line indicates that it also traces the outflow. This is much clearer in the [Ne\,III]~15.6~$\mu$m line, which has a higher IP of 41.0~eV and is therefore more influenced by AGN photoionisation. 
And in both the [Ne\,V]~24.3~$\mu$m and [Ne\,VI]~7.7~$\mu$m lines, with IP of 97.1~eV and 126.2~eV respectively, the emission only has high enough surface brightness to be spatially resolved in the direction of the outflow. In both these lines, the kinematics are completely dominated by the outflow.
For the [Ne\,V] line, the high dispersion oriented perpendicular to the outflow direction is likely to be a beam smearing effect due to the broader PSF ($\sim$0.9\arcsec\ FWHM) and the rapid change from highly red shifted to highly blue shifted velocities.
This effect is not so prominent in the [Ne\,VI] line where the PSF has a FWHM of $\sim$0.4\arcsec.
A different feature that is very clear in the kinematics maps of the [Ne\,III] line is the anomalous velocity and high dispersion extending from the nucleus along the minor axis out to 2--3\arcsec\ where the surface brightness of the line emission is much lower.
Such features have been reported for other galaxies \citep{ven21,per23,aud23}, where the explanation is based on models of young jets in disks \citep{muk18,mee22}. This is understood to be lateral turbulence caused by a jet that is inclined towards the host galaxy disk. As it is decelerated and deflected by clouds in the disk, it drives a sub-relativistic wide-angled outflow normal to the disk plane. The shocks caused by this jet-driven bubble increase the velocity dispersion of gas in directions away from the jet itself.
It has been identified in a number of other galaxies in our full sample as described in Zhang et al. (in prep).

Maps of the flux and kinematics extracted for several H$_2$ lines are shown in Fig.~\ref{fig:h2_maps}. 
In addition to the velocity and velocity dispersion, the figure shows the residual velocity after subtracting the velocity field of the disk model fitted to the CO\,(2-1) kinematics in \cite{shi19}.
These data are analysed in detail in the following sections, and so here we note only the most striking features.
The first of these is that the morphology and kinematics of the CO and S(1) lines are remarkably similar indicating they trace the same molecular gas component.
The second is that, while CO is not detected in the ionisation cones, the rotational H$_2$ lines are detected. This confirms that molecular gas can survive in the outflow.
The velocity field and its residual show that the molecular gas in the disk is perturbed as it passes through the outflow, an effect that is more prominent in the higher excitation molecular lines.
It is these features that motivate the analysis of the H$_2$ excitation and temperature distribution in Sec.~\ref{sec:h2temp} and of the H$_2$ kinematics in Sec.~\ref{sec:h2_kin}.

\begin{figure*}
    \centering
    \includegraphics[width=13cm]{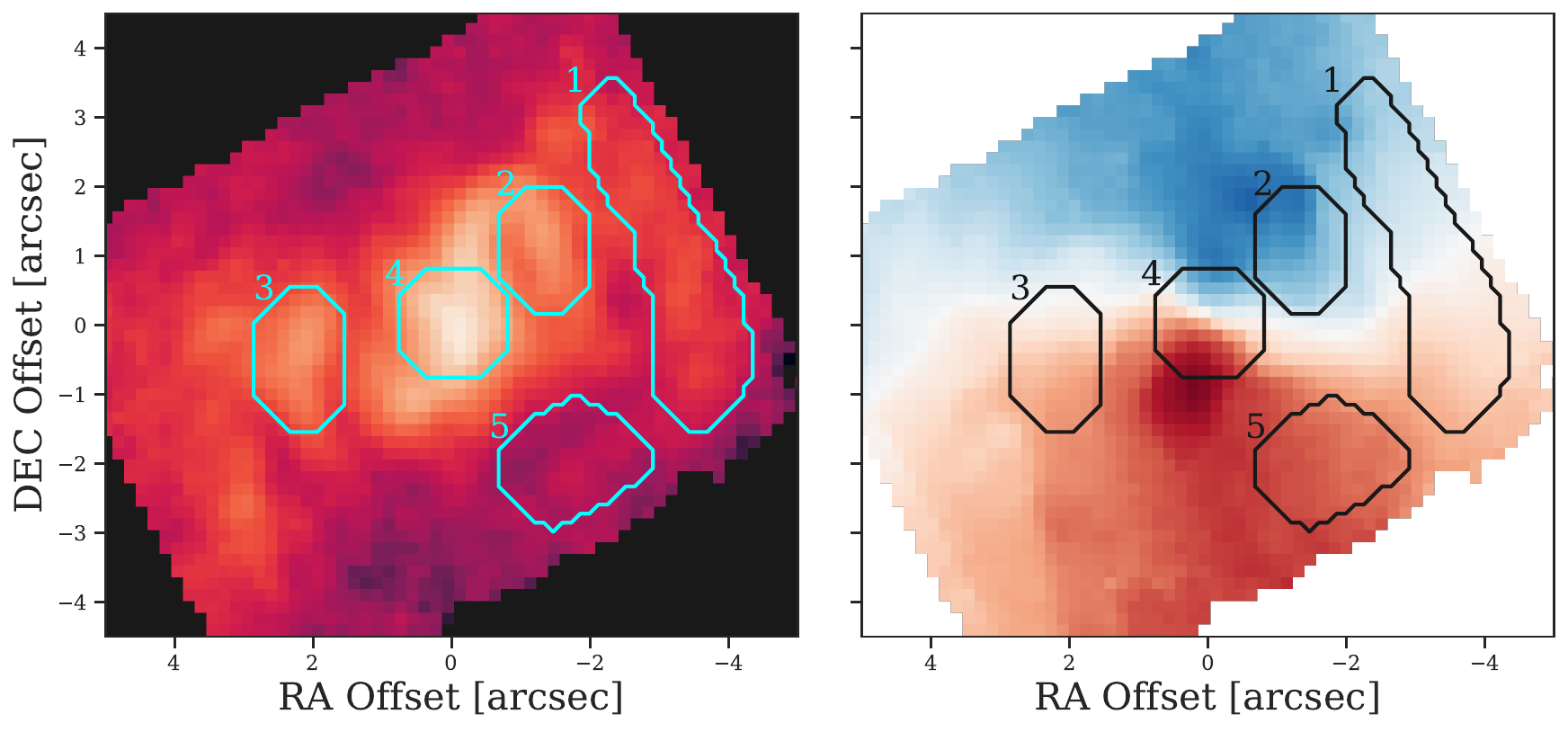}
    \caption{Apertures superimposed on the S(5) flux map. Rather than being circular, these trace specific features within the overlapping fields of view: (1) the star forming arc, (2) an H$_2$ hotspot in the ionisation cone, (3) a knot of H$_2$ emission to the east, (4) the nucleus, and (5) a region to the south outside the ionisation cone.}
    \label{fig:h2aps}
\end{figure*}

\begin{figure*}
    \centering
    \includegraphics[width=\textwidth]{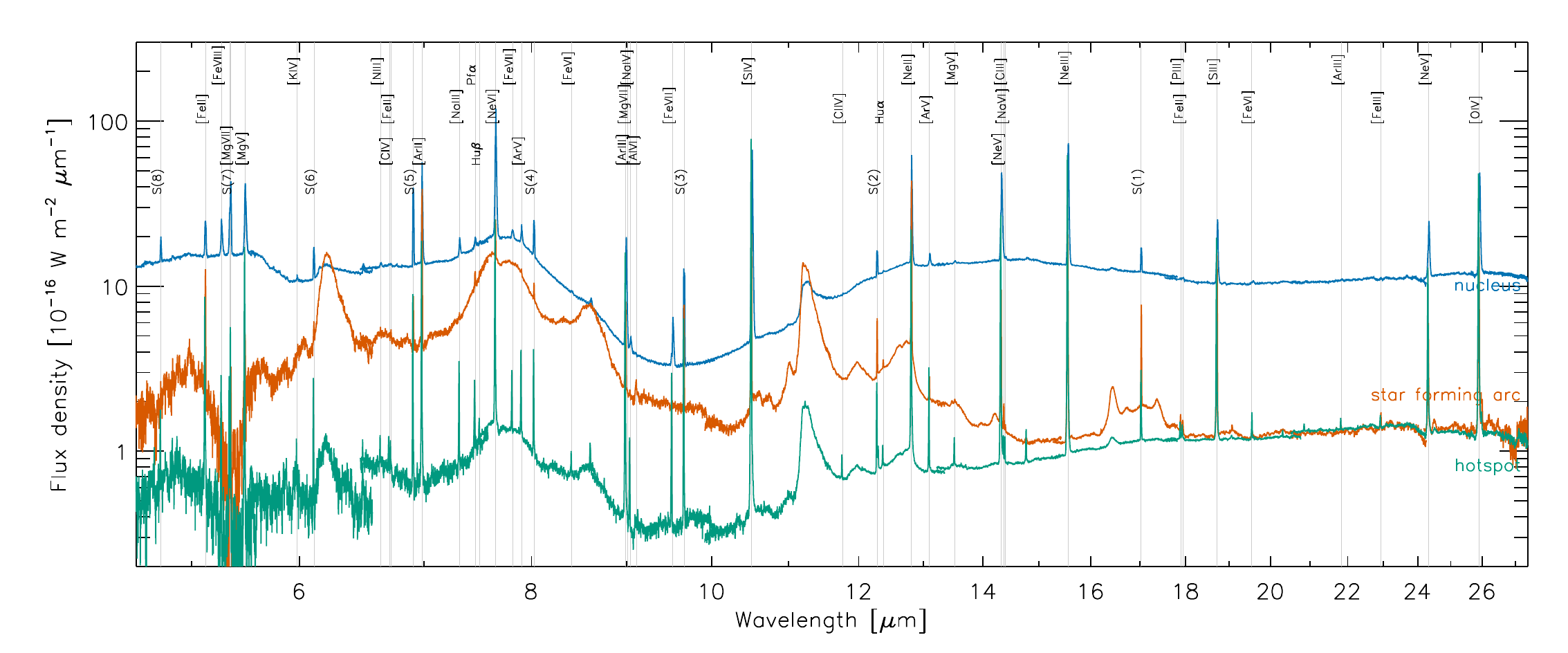}
    \caption{Mid-infrared spectra in NGC\,5728. The spectra for three apertures, corresponding to the nucleus (blue; region 4 in Fig.~\ref{fig:h2aps}), star forming arc (red; region 1 in Fig.~\ref{fig:h2aps}), and hotspot (green; region 2 in Fig.~\ref{fig:h2aps}) are shown on a logarithmic flux scale. The spectra have been extracted from each of the 12 separate datacubes and overplotted. They show that no re-scaling between the cubes is needed. The most prominent lines with reliable identifications have been marked: 36~fine structure lines, 3~H\,I lines, and 8~H$_2$ lines.}
    \label{fig:spec}
\end{figure*}

\section{Molecular gas excitation and temperature distribution}
\label{sec:h2temp}

The MRS bandpass covers the rotational H$_2$ transitions from 0-0\,S(8) at 5.05\,$\mu$m to 0-0\,S(1) at 17.04\,$\mu$m (the 0-0\,S(0) line is outside the currently available wavelength range).
These have excitation temperatures down to 1015\,K and are able to probe molecular gas as cool as $\sim$100\,K, making them an ideal complement to the transitions accessible to millimetre interferometers that tend to probe the coldest component of the gas.
In addition, observing eight transitions makes it possible to assess the spatially resolved temperature distribution of the gas.
In order to mitigate the impact of the variation in spatial resolution over the wide wavelength range, after resampling the individual cubes to a common pixel grid of 0.13\arcsec\ (corresponding to the smallest pixel scale of the full data set), we extracted line fluxes in apertures with a minimum width of 0.8\arcsec\ and mostly $>$1\arcsec. Tests indicate that in this case PSF effects have no more than a 20\% impact on extracted fluxes from S(1) to S(8).
This upper limit to the systematic error matches the 20\% statistical uncertainty that we consider sufficient for distinguishing models of the H$_2$ population distribution at a meaningful level. We note that it would be incorrect to apply a PSF correction because of the extended nature of the spatial distributions of the lines.
The apertures, which have areas in the range 1.7--5.0~sq.~arcsec, are irregular in shape because they trace specific features on the emission line maps that characterise the different environments in the field of view. As shown in Fig.~\ref{fig:h2aps}, these correspond to (1) the star-forming arc, (2) a hotspot in the ionisation cone to the north-west of the nucleus (P5 in \citealt{dur18}), (3) a knot of emission to the east of the nucleus, (4) the nucleus itself (P1 in \citealt{dur18}), and (5) a region to the south that is outside the ionisation cone. The spectra extracted in three of the apertures are shown in Fig.~\ref{fig:spec}, and the H$_2$ line fluxes for all five apertures are reported in Table~\ref{tab:h2mrs}.

The analysis we perform in this Section encompasses several aspects, and so we provide a brief summary here before going into the details.

We first look at the populations of the $J=3$--10 levels in all the apertures. These show two distinct groupings, hinting at minimum temperatures of 300-400\,K for the $J=3$--5 levels; but clear evidence for warmer gas that dominates the higher-$J$ transitions. We make simple models of these distributions assuming local thermal equilibrium (LTE) in order to assess the temperature distribution in the different regions. The ortho-para ratio in the LTE models provides a good match to the data, indicating that such models are appropriate for the pure rotational levels (although we find later that the vibrational levels show evidence for an ortho-para ratio < 3, indicating they are modified by non-thermal excitation).

Based on these results, we create an excitation map by looking at the ratio of the S(1) and S(5) lines, since the corresponding $J=3$ and $J=7$ levels are dominated by cooler and warmer gas respectively. This provides more insights into the spatial distribution, in particular with respect to the boundaries of the ionisation cone near the circumnuclear disk.

The remaining analysis focuses on the nuclear aperture, for which we include $v$=1-0 and $v$=2-1 ro-vibrational transitions from the near-infrared data \citep{shi19}. 
In addition, comparing the Br$\gamma$ and Pf$\alpha$ lines at 2.17~and 7.46\,$\mu$m provides an estimate of the extinction towards the nuclear emission (see \citealt{don24} for a quantitative assessment of the merits of extinction tracers and models).  This has modest impact on the rotational lines except to correct S(3) for the silicate absorption; but is important here because of the differential effect between the mid-infrared and near-infrared regimes.
The additional near-infrared H$_2$ transitions show that the models we have used for the pure rotational lines at moderate temperatures cannot be directly extended to higher temperatures. Instead, we use a non-LTE model because of the greater critical densities for those lines, and include H because their critical densities for collisions with H are less than for collisions with H$_2$ (see \citealt{leb99}). As such, the model is also able to provide information about the local gas volume density. But while this model is able to indicate gas conditions that can give rise to the observed line emission, it does not provide a physical mechanism to achieve it. We therefore turn to a library of shock models to assess whether this is a plausible and likely mechanism.

\begin{table*}
\caption{H$_2$ 0-0 line fluxes measured in the five apertures shown in Fig.~\ref{fig:h2aps}}
\label{tab:h2mrs}
    \begin{tabular}{llccccc}
        \hline\hline
        Line & $\lambda$ ($\mu$m) & \multicolumn{5}{c}{Flux ($10^{-18}$~W~m$^{-2}$)} \\
        & & star forming arc & hotspot & eastern knot & nucleus & southern region \\
        & & (5.0~sq.\,\arcsec) & (1.7~sq.\,\arcsec) & (2.0~sq.\,\arcsec) & (1.8~sq.\,\arcsec) & (2.5~sq.\,\arcsec) \\
        \hline
        0-0 S(1) & 17.035 & $7.70\pm0.24$ & $2.44\pm0.05$ & $4.99\pm0.08$ &  $12.30\pm0.23$ & $2.35\pm0.03$ \\
        0-0 S(2) & 12.279 & $3.41\pm0.12$ & $1.74\pm0.03$ & $2.20\pm0.04$ & $~~9.36\pm0.17$ & $0.84\pm0.04$ \\
        0-0 S(3) & 9.6649 & $4.55\pm0.19$ & $4.00\pm0.09$ & $4.12\pm0.10$ &  $15.60\pm0.25$ & $0.95\pm0.05$ \\
        0-0 S(4) & 8.0251 & $1.48\pm0.16$ & $1.69\pm0.05$ & $1.11\pm0.06$ &  $12.00\pm0.17$ & $0.30\pm0.05$ \\
        0-0 S(5) & 6.9095 & $2.41\pm0.16$ & $3.78\pm0.06$ & $1.92\pm0.08$ &  $26.20\pm0.31$ & $0.50\pm0.07$ \\
        0-0 S(6) & 6.1086 & $0.76\pm0.15$ & $0.88\pm0.07$ & $0.32\pm0.07$ & $~~5.63\pm0.10$ &       $<0.27$ \\
        0-0 S(7) & 5.5112 & $0.86\pm0.37$ & $1.78\pm0.14$ & $0.67\pm0.13$ &  $21.70\pm0.37$ &       $<0.51$ \\
        0-0 S(8) & 5.0530 &       $<0.99$ & $0.37\pm0.11$ &       $<0.45$ & $~~3.48\pm0.15$ &       $<0.54$ \\
        \hline
    \end{tabular}
\end{table*}

\begin{figure}
    \centering
    \includegraphics[width=9cm]{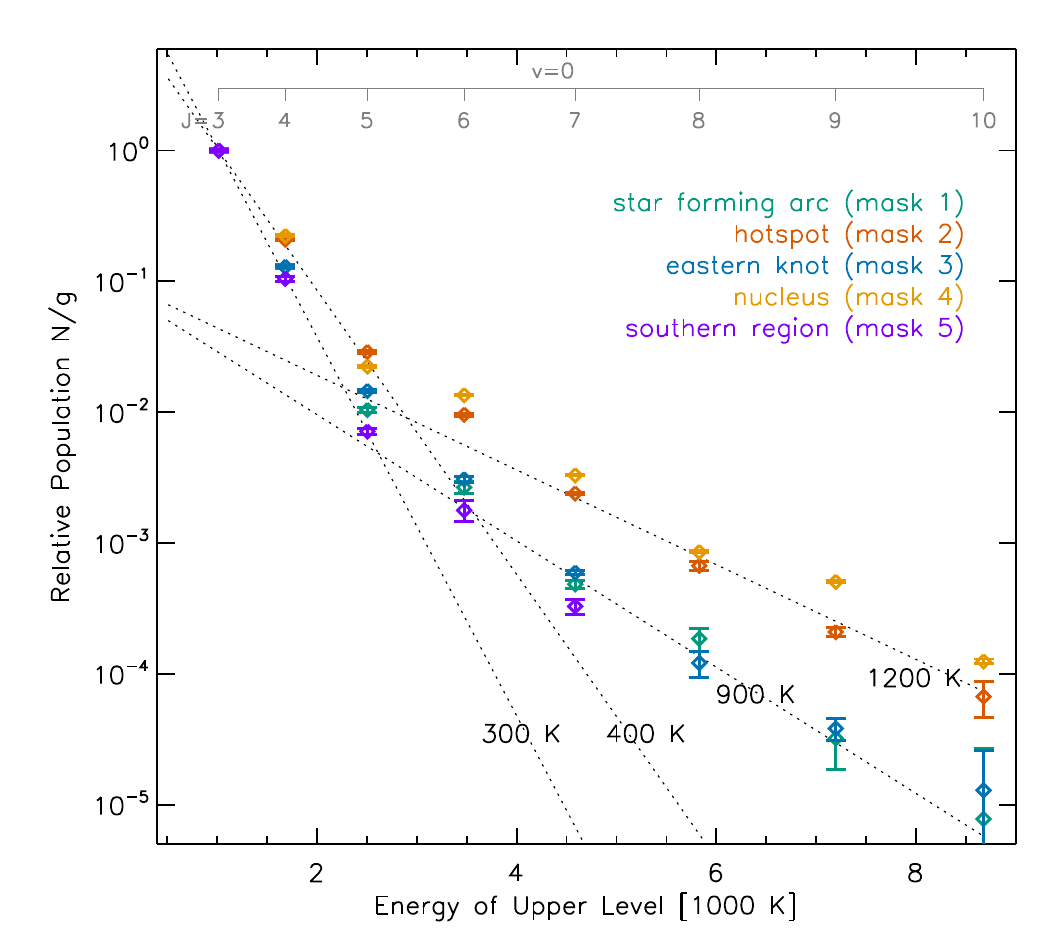}
    \caption{Population levels for several apertures in the field of view. The level populations derived from their extracted fluxes showing that these fall into  two groups. The dashed lines for single temperatures of 300~K, 400~K, 900~K and 1200~K characterise the temperature range probed by the lines.}
    \label{fig:h2pops}
\end{figure}

\begin{figure}
    \centering
    \includegraphics[width=9cm]{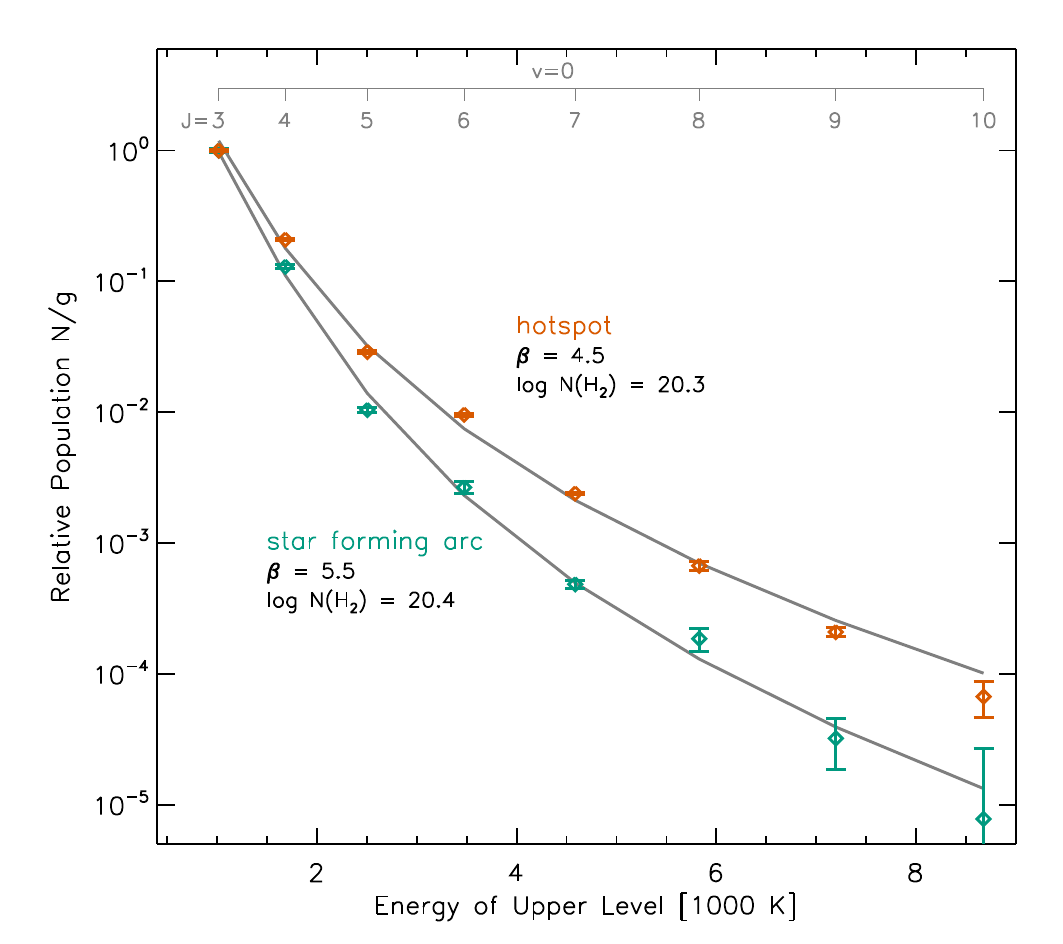}
    \caption{Fits for a power-law temperature distribution. Fits to the star forming arc in the circumnuclear ring and the hotspot in the ionisation cone, show they can be well characterised by such a distribution.}
    \label{fig:h2plfit}
\end{figure}

\begin{table}
\caption{Parameters from LTE models of the H$_2$ line ratios}
\label{tab:ltefit}
    \begin{tabular}{lccc}
        \hline\hline
        Aperture & $\beta$ & log $N_{H_2}$ (cm$^{-2}$) & log M ($M_\odot$) \\
                 & index   & \multicolumn{2}{c}{at $T > 200$ K} \\
        \hline
        star forming arc    & 5.5 & 20.4 & 5.8 \\
        hotspot             & 4.5 & 20.3 & 5.3 \\
        eastern knot        & 5.5 & 20.6 & 5.7 \\
        nucleus             & 4.4 & 21.2 & 6.3 \\
        southern region     & 6.0 & 20.2 & 5.3 \\
        \hline
    \end{tabular}
\end{table}

\subsection{Thermal equilibrium models}
\label{subsec:thermal}

We have extracted fluxes of the S(1) to S(8) lines in the apertures indicated in Fig.~\ref{fig:h2aps}.
The derived population levels $N/g$ are shown in Fig.~\ref{fig:h2pops} as a function of the energy of the level (as a temperature). 
These fall into two distinct groupings, neither of which can be characterised by a single temperature.
The populations for the star forming arc, eastern knot, and southern region have similar distributions, and indicate the gas is relatively warm, with temperatures in the range 300-900\,K.
The nucleus and hotspot are warmer still, in the range 400--1200\,K.

Although the data can be characterised by two temperatures, a more physical approach is to adopt a power-law temperature distribution  
d$N = T^{-\beta}$d$T$, where $N$ is the gas column density \citep{zak10,per14,tog16}. 
Applying this method yields the fits in Fig.~\ref{fig:h2plfit}.
In all cases the $J=5$ level is below the fit, but that is because the corresponding S(3) line at 9.66\,$\mu$m is affected by the silicate absorption. 
The result here is that the star forming arc, eastern knot, and southern region have power-law indices in the range $\beta = $5.5--6.0, while the nucleus and hotspot have significantly shallower indices of $\beta = $4.4--4.5 indicating a larger fraction of warmer gas.
Considering the contributions of gas at different temperatures to each level, as illustrated in Fig.~\ref{fig:h2pops}, these models indicate that the S(1) line flux is dominated by gas at temperatures of 200~K or slightly less while S(7) is dominated by gas at 1400-1600~K.
Since the lines we observe are insensitive to gas at temperatures below $\sim200$~K, care needs to be taken if extrapolating the power-law index to lower temperatures. Indeed, by simultaneously fitting CO and H$_2$ line distributions, \citet{per14} have shown that, at least in Seyfert LIRGs, a flatter index is appropriate at $T<200$\,K.

The initial studies adopting a power-law index to characterise the rotational H$_2$ lines focused on ultra-luminous infrared galaxies (ULIRGs). 
\citet{zak10} found indices in the range $\beta=2.5$--5.0 could reproduce the variety of observed excitation diagrams, while \citet{per14} derived values consistent with only the top end of that range.
The sample of \citet{tog16} was largely based on the Spitzer Infrared nearby Galaxies Survey \citep{ken03} and so contained a wider variety of objects (LINERs and Seyferts, as well as H{\sc II} and dwarf galaxies) to which some radio galaxies and a quasar were added. They found values for $\beta$ in the range 4--6, with an average of $\beta=4.8\pm0.6$ for the SINGS galaxies.
More recently, in the circumnuclear region of NGC\,7469, \citet{zha23} found values varying from $5$ at a radius of 600~pc to $4$ at the nucleus.
The range of $\beta$ values we find in all five apertures are consistent with these and may be associated with a shock origin for the heating mechanism. This was explicitly addressed by \citet{per14}, who applied the same analysis to models of line emission calculated for gas that cools radiatively after it has been heated at a constant rate to a temperature that is dependent on the adopted shock velocity. For gas densities n$_{H_2}\gtrsim10^5$\,cm$^{-3}$ they found $\beta\gtrsim4$. In contrast, they found that PDR models yielded rather high values of $\beta>10$. Nevertheless, they argued that PDR emission may be needed to account for the CO emission line ratios, and that a combination of the two processes was likely. 
We will return to the topic of shock excitation, and take a more detailed look at it, in Sec.~\ref{subsec:shocks}.

\begin{figure}
    \centering
    \includegraphics[width=\columnwidth]{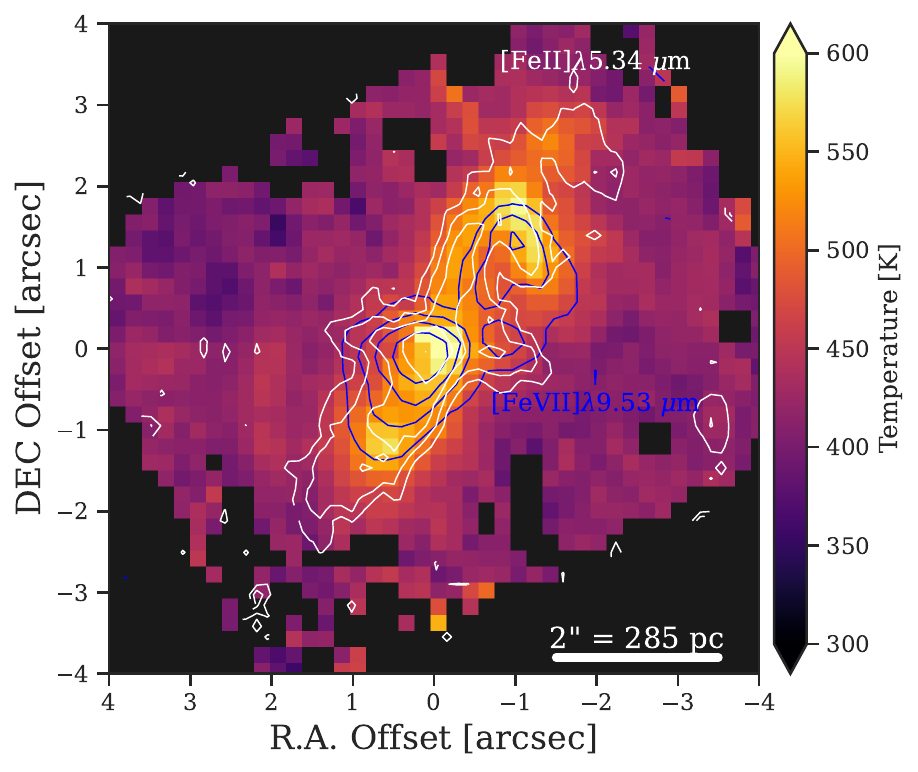}
    \caption{H$_2$ excitation map. 
    Map of molecular gas temperature derived from the S(1)/S(5) line ratio (shown at the 0.20\arcsec\ pixel scale of the S(1) line). This characteristic temperature also represents the distribution of temperatures at each location. 
    The comparison to the [Fe\,II] and [Fe\,VII] lines (white and blue contours respectively) show that the warmer H$_2$ emission traces the edge rather than bulk of the ionisation cone. Specifically, it traces the northern edge of the western cone and the southern edge of the eastern cone.}
    \label{fig:h2_temp}
\end{figure}

\subsection{Excitation distribution: jet or disk interaction?}
\label{subsec:tmap}

Looking at the spatially resolved H$_2$ excitation distribution can help to assess whether it is heated by an interaction of the disk with the jet or the ionisation cone.  
The integral field capability of the MRS enables us to do so in a simple but robust way, deriving the excitation from the ratio of a pair of H$_2$ lines. Because the S(3) line is close to 10~$\mu$m and hence affected, particularly in the very centre, by silicate absorption (see Sec.~\ref{subsec:nonlte}), we have used the ratio of the S(1) and S(5) lines. 
The analysis above shows that this can be considered as a characteristic temperature, or equivalently as indicative of the index for a power-law distribution at each location.
Adopting the former representation, the temperature map we derive is shown in Fig.~\ref{fig:h2_temp}.

It is difficult to make conclusive statements about whether this is the result of an interaction between the jet and disk. 
In favour of such an explanation is the brightening of the jet at a distance of about 2.5\arcsec\ from the core (Fig.~\ref{fig:h2_maps}), which is close to the hotspot at a radius of 1.8\arcsec\ (region 2 in Fig.~\ref{fig:h2aps}), suggesting a possible link between these phenomena.
However, it is difficult to associate the jet with the remarkable coincidence between the location of the warmest H$_2$ and the edge of the ionisation cone that is revealed by Fig.~\ref{fig:h2_temp}. Here, the high excitation [Fe{\sc vii}] 9.53\,$\mu$m line (blue contours) traces the bulk of the ionisation cone, while the low excitation [Fe{\sc ii}] 5.34\,$\mu$m line (white contours) traces an X-shape corresponding to the partially ionised zone at its edges.
The excitation map very clearly follows the edge of the ionisation cone, but only on one side: to the west it matches the northern edge but is absent on the southern edge, while to the east it matches the southern edge but is absent on the northern edge.
As discussed by \citet{shi19}, the disk of NGC\,5728 rotates in a clockwise direction. And hence, on the western side of the disk, gas enters the ionisation cone from the north. Similarly, on the eastern side of the disk, gas rotates into the ionisation cone from the south.
The H$_2$ temperature map is tracing these specific boundaries.
The spatial coincidence with the [Fe{\sc II}] emission along these edges suggests that the molecular gas in the disk is heated as it enters the ionisation cone, as the clouds in the disk collide with gas that is being accelerated outwards by the AGN.

Both of the processes discussed above could be a source of shocks, which would be consistent with the expectation from Sec.~\ref{subsec:thermal}.
And while we can only use qualitative arguments, it seems plausible that both are involved: the jet-disk interaction creating the hotspot to the north-west of the AGN, and the rotation of disk gas into the ionisation cone to explain the asymmetric one-sided heating.

\begin{table}
\caption{Additional line fluxes measured in the nuclear aperture}
\label{tab:h2sinf}
    \begin{tabular}{lcc}
        \hline\hline
        Line & $\lambda$ ($\mu$m) & Flux ($10^{-18}$ W\,m$^{-2}$)\\
        \hline
        Br$\gamma$ & 2.1661 & $1.82\pm0.02$ \\
        Pf$\alpha$ & 7.4599 & $2.64\pm0.17$ \\
        1-0 S(0) & 2.2235 & $1.42\pm0.03$ \\
        1-0 S(1) & 2.1218 & $5.54\pm0.09$ \\
        1-0 S(2) & 2.0338 & $2.01\pm0.03$ \\
        1-0 S(3) & 1.9576 & $5.22\pm0.08$ \\
        2-1 S(1) & 2.2477 & $0.69\pm0.01$ \\
        2-1 S(2) & 2.1542 & $0.39\pm0.02$ \\
        2-1 S(3) & 2.0735 & $0.73\pm0.02$ \\
        \hline
    \end{tabular}
\end{table}

\begin{figure}
    \centering
    \includegraphics[width=\columnwidth]{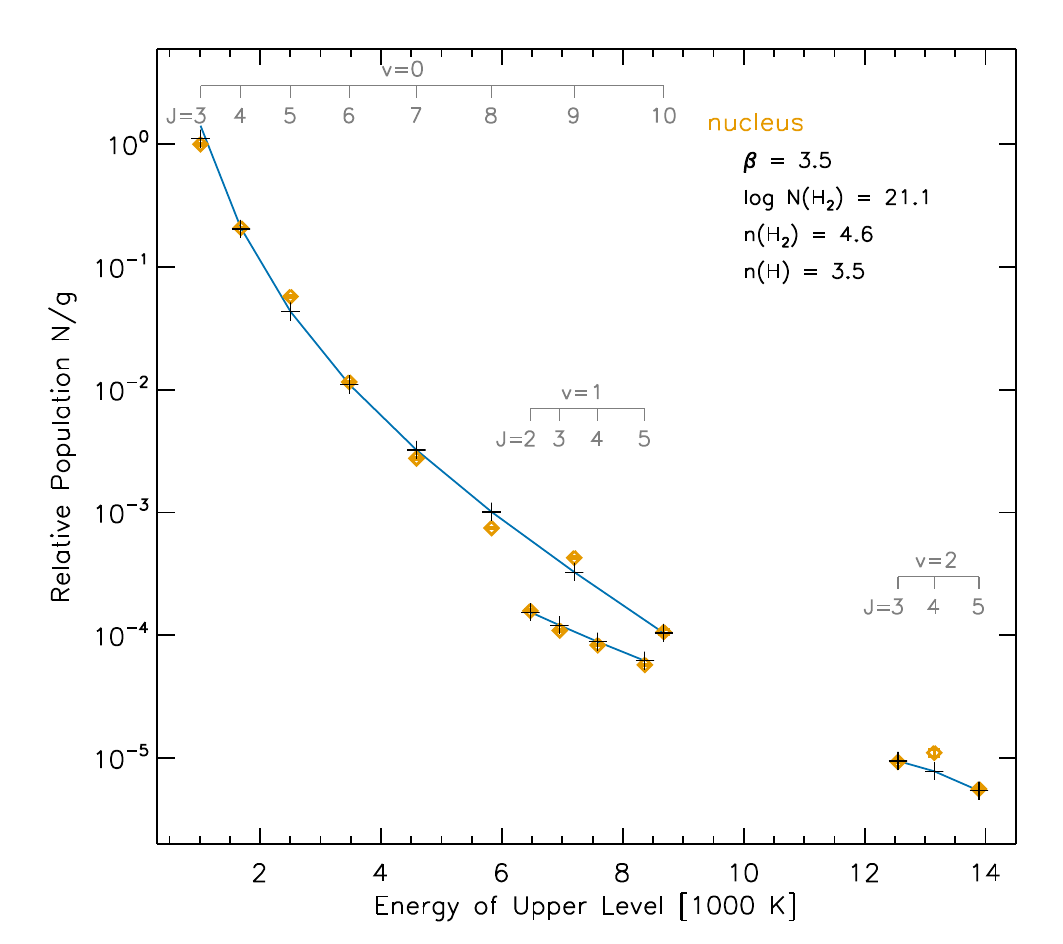}
    \caption{Non-LTE fit with a power-law temperature distribution, to the nuclear H$_2$ populations derived from the rotational and vibrational lines, after correcting for extinction. In this case, the volume density of both H$_2$ and H are derived because of their influence on collisional de-excitation.}
    \label{fig:nonltefit}
\end{figure}

\subsection{Non-LTE model of the nuclear region}
\label{subsec:nonlte}

We can extend the analysis of the nucleus itself by including seven ro-vibrational $v$=1-0 and $v$=2-1 observed in the K-band with SINFONI, given in Table~\ref{tab:h2sinf}.
These were measured from the reduced cubes analysed by \citet{shi19}, who only included the 1-0\,S(1) line in their work.
Estimating the extinction to the line emitting region is important here because of its differential effect between the near-infrared and mid-infrared lines.
We therefore include the Br$\gamma$ line from SINFONI and the Pf$\alpha$ line from the MRS spectrum.
We have considered three prescriptions for the extinction curve: \citet{ros00} towards the Orion Molecular Cloud, \citet{chi06} for the local ISM, and \citet{fri11} towards the Galactic Center.
The overall shapes of these curves across the 2-20~$\mu$m range are similar, the major difference being their treatment at 8--12~$\mu$m where silicate absorption is prominent, and which in our analysis only affects the S(3) line.
Trying these different curves yields consistent fits to the extinction corrected data, well within the uncertainties from the measurements themselves, and so for the discussion here we simply adopt that of \citet{fri11}.
From the ratio of the Br$\gamma$ and Pf$\alpha$ lines we find $A_{Br\gamma} = 0.78$\,mag.
The impact is relatively modest in the infrared, although it would imply the nucleus is not visible at optical wavelengths.
The main effect is in the silicate absorption band, modifying the ratio of S(3)/S(1) by a factor 2.6, and the differential reddening, which is a factor 1.4 between the near- and mid-infrared from 1-0\,S(1) to 0-0\,S(1). 

The full set of H$_2$ transitions cannot be fit with an LTE model that has a single power-law temperature distribution. We therefore drop the assumption of LTE.
Doing so is physically motivated because the ro-vibrational lines have much higher critical densities and hence are not necessarily thermalised. Indeed, observations of AGN show evidence of non-thermal line ratios in the $v$=2-1 transitions \citep{dav05}.
As such, the volume density of H$_2$ becomes a parameter of the model.
In addition, because the critical densities for these transitions are lower by 2-3 orders of magnitude when considering H as the perturber rather than H$_2$ \citep{leb99}, the density of the neutral component must also be included.
The model we adopt here is described in \citet{per14} which makes use of RADEX \citep{van07}. The grids considered cover kinetic temperatures $T_{kin} = 10$--2800\,K, H$_2$ density n$_{H_2} = 10^2$--10$^9$\,cm$^{-3}$, ratio n$_H$/n$_{H_2} = 1$--10$^{-5}$. The ortho-to-para ratio adopted varies between 0 for low temperatures and 3 for T$_{kin}>200$\,K (see \citealt{bur92}). These models are then used with a power-law temperature distribution as described above.

Fig.~\ref{fig:nonltefit} shows that the full set of transitions can be fit with a single power law temperature distribution. 
In this case we find $\beta = 3.6$, slightly flatter than, but consistent with, the result for the LTE fit to the rotational transitions.
The densities yielded by the fit, of n$_{H_2} = 10^{4.5}$\,cm$^{-2}$ and n$_{H} = 10^{3.4}$\,cm$^{-2}$, are needed to match the relative populations between the vibrational levels in the absence of any external influence.
However, the figure also shows that the $v$=1-0 transitions are thermalised, while there are indications for non-thermal processes in the $v$=2-1 transitions: the J=4 level is high with respect to J=3 and J=5 as expected for an ortho-para-ratio $<3$.
This model would need an additional component -- such as an external UV field -- to achieve that.
Similar effects have been reported before for AGN, where they were interpreted as due to a dense PDR \citep{dav05}.
Here, the power-law index and densities are consistent with the calculation performed by \citet{per14} for shocked gas.
We therefore explore whether shocks, perhaps illuminated by an external UV field, may be the origin of the line emission.

\begin{figure}
    \centering
    \includegraphics[width=\columnwidth]{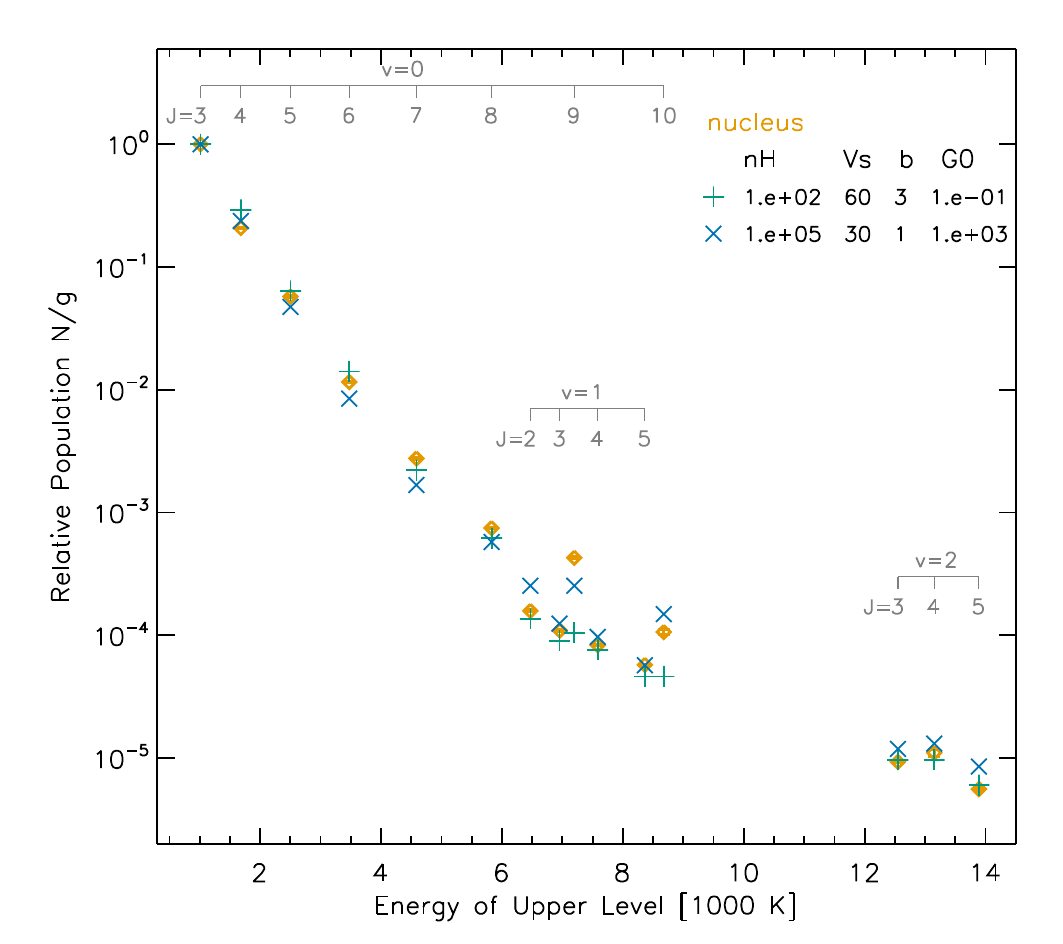}
    \caption{Two potential shock models from \citet{kri23} that are a reasonable match to the nuclear H$_2$ populations (after correcting for extinction). Their main parameters (density n$_H$, velocity V${\rm s}$, magnetic field scaling factor $b$, and radiation field G$_0$) are indicated.}
    \label{fig:shockmodel}
\end{figure}

\subsection{Shock models of the nuclear region}
\label{subsec:shocks}

In order to assess whether shock models are consistent with the observed line emission, we use the library of models calculated by \citet{kri23}, which are based on models of \citet{flo85} and \citet{god19}.
These authors note that the rotational H$_2$ lines alone are unlikely to strongly constrain the grid of models, and so we have limited this analysis to the nuclear region (Fig.~\ref{fig:h2aps}) where we can combine the $v$=0-0, $v$=1-0 and $v$=2-1 transitions.
In addition, rather than focus solely on the model with minimum $\chi^2$, we consider the properties of all of the models with lower $\chi^2$.
Among the matches with the lowest $\chi^2$ are two groups of models, an example of each of which is shown in Fig.~\ref{fig:shockmodel}.
Each of these groups points to a characteristic density n$_H$, shock speed V${\rm s}$, magnetic field strength $b$ and radiation intensity $G_0$, while the cosmic ionisation rate $\zeta$ and PAH abundance are unconstrained.

One model group, denoted by green plus signs, represents a higher velocity V${\rm s} = 60$~km\,s$^{-1}$ shock in a lower density log~n$_H$~(cm$^{-3}$) = 2 medium (note that V${\rm s}$ represents the shock speed internal to the clouds and is not necessarily indicative of the bulk cloud speed). 
This remains a C-shock because the magnetic field scaling factor $b=3$ is relatively high (the magnetic field $B$ in the models is transverse, given by $b\sqrt{n_H}$, here corresponding to 30~$\mu$G).
Only a weak interstellar radiation field $G_0 = 0.1$ is required because the density is too low to thermalize the 2-1 levels. 
This model matches the ro-vibrational levels very well, and also the lower rotational levels. However, it fails for the $J=9$ and $J=10$ levels, under-predicting these levels by a factor 2--4. This indicates that the temperature distribution of the gas is too steep; and if one were able to measure the populations of higher rotational levels these would probably be even more significantly under-estimated by the model. For this reason we consider this model less likely.

The other model group, denoted by blue crosses, represents lower velocity V${\rm s} = 30$~km\,s$^{-1}$ shock in a higher density log~n$_H$~(cm$^{-3}$) = 5 medium.
This higher density matches the expectation from our previous analysis much better.
While the magnetic field scaling parameter $b = 1$ is lower, the higher density means the actual magnetic field is also higher at $B = 300~\mu$G. 
Finally, this group of models has a very significant incident radiation field $G_0 = 10^3$, which is the cause of the non-thermal signature of the $v$=2-1 levels. So close to the AGN, the origin of the high radiation field is expected to be the AGN itself. 
For this model, the typical difference to the data is a factor 1.3, with a maximum of only 1.7, and so we consider it more representative of the full data set.
That the fit is not better is a combination of the coarse spacing in the model grid and that within the aperture of area of $\sim$0.1~kpc$^2$ it is likely more than one excitation process is at work.

While these models appear to support the contention that the disk gas has been shock heated (e.g. by the jet or as it rotates into the ionisation cone), we need to assess whether the H$_2$ line luminosities are consistent with that of CO\,(2-1) or whether there is additional cold molecular gas that is not shock heated. This can be done by a comparison of their respective column densities.

\subsection{Gas Column Density}
\label{subsec:column}

The LTE fits discussed in Sec.~\ref{subsec:thermal} indicate column densities (2H$_2$ + H) in the range $\log{N_H~(cm^{-2})} = 20.5$--21.5, where the highest value is for the nuclear aperture.
The column inferred from the more detailed non-LTE analysis of the nucleus in Sec.~\ref{subsec:nonlte} is consistent with this at $\log{N_H} = 21.4$.
However, a caveat for these estimates is that the models were truncated at a lowest temperature of 200~K, meaning that any substantial gas component at a significantly cooler temperature is not included.
This is important to bear in mind when comparing the column densities to those implied by the shock models discussed in Sec.~\ref{subsec:shocks}.
Our preferred model yields $\log{N_H} = 21.3$, although the relative contribution of H and H$_2$ to this total is rather different. We assign this to the impact of the external UV field in the shock model, which helps to adjust the relative vibrational levels without needing to rely solely on collisional processes.
Since the shock model integrates the full gas column, this similarity means that within the nuclear aperture the bulk of the gas in the model must be at $\gtrsim$200~K -- a situation that should be expected since the post-shock temperature is $>200$~K.
We can confirm this by considering the $J<3$ populations in the shock model.
These are within 10\% of an isothermal line with a temperature of about 370~K, indicating that, despite their lower excitation energies, they do not trace cooler gas components than the $J=3$ population.

Given the similarity of the H$_2$ rotational populations between the nucleus and the hotspot, we apply the same shock model to the hotspot and compare the column densities. 
Based on the fit for a power-law temperature distribution in LTE, we find $\log{N_H} = 20.6$ (Table~\ref{tab:ltefit}), consistent with the $\log{N_H} = 20.6$ obtained by scaling the shock model to the observed flux.
Thus the simple LTE analysis of this extra nuclear region is consistent with the same shock model that matches the more detailed data available for the nucleus.

We now turn to the CO\,(2-1) emission reported in \cite{shi19}.
Those authors adopted a CO-to-H$_2$ ratio of $\alpha_{CO}=1.1$ typical for the central regions of nearby galaxies.
However, in a detailed study of NGC\,3351, \cite{ten22} found that while this is what one might expect when integrating over the whole central region, $\alpha_{CO}$ shows substantial variations from region to region.
In particular, they found that in turbulent flows, the average could be an order of magnitude lower.
They explained this in terms of the higher velocity dispersion leading to lower optical depths and a larger fraction of the CO emission being able to escape.
Although CO\,(2-1) is not detected in the outflow in a spatially resolved sense, integrating over the hotspot aperture provides a marginal detection that would yield a column density similar to that above if one adopts $\alpha_{CO}\sim0.3$.
Low values comparable to this have been reported by other authors, such as \citet{mps24} who derived $\alpha_{CO}\sim0.4-0.6$ in the molecular outflow of NGC\,3256 using the CO $v$=1-0 4.67~$\mu$m band observed with NIRSpec, and \citet{ten22} who found $\alpha_{CO}<0.1$ in turbulent inflowing gas in NGC\,3351 based on multiple CO, $^{13}$CO and C$^{18}$O transitions.
In our case, it would imply that all the molecular gas in the outflow is heated by the shock, and no cold component remains. As such, the CO\,(2-1) line is weak but one might expect to detect higher transitions.
However, in the nuclear aperture, using the same value of $\alpha_{CO}$ yields $\log N_H = 22.3$, which is an order of magnitude higher than indicated by the analysis of the H$_2$ rotational lines.
This strongly suggests there is an additional cold component in the nucleus.
This is apparent in the CO\,(2-1) map in Fig.~\ref{fig:h2_maps} as a transverse component, which was identified as {\it Comp2} by \cite{shi19}. They described it as a warped inner rotating molecular disc that contributes to shaping the bicone and provides the material fuelling the AGN, and is associated with the higher extinction at the nucleus.
More recent data with beam sizes of 0.2--0.4\arcsec\ in the CO\,(1-0), CO\,(2-1), and CO\,(3-2) transitions spatially resolve this highly inclined disc, and indicate that the typical physical conditions correspond to $n_H \sim10^2-10^5$~cm$^{-3}$ and $T_{kin}\sim10-50$~K (Garc\'ia-Burillo et al. in prep).
Thus, in contrast to the hotspot region, in the nuclear aperture, the shock excited H$_2$ rotational lines do not trace the full gas column.

Our conclusion is that the available evidence points to a scenario in which the molecular gas in the ionisation cone is excited both by shocks and UV radiation from the AGN. 
Because the bulk of the molecular gas in the outflow is shock heated, the CO\,(2-1) line is no longer a good tracer.
In the central arcsec, while the H$_2$ rotational lines still trace the shocked gas, the CO line emission is strong because it is dominated by a separate highly inclined disk.

\begin{figure*}
    \centering
    \includegraphics[width=\textwidth]{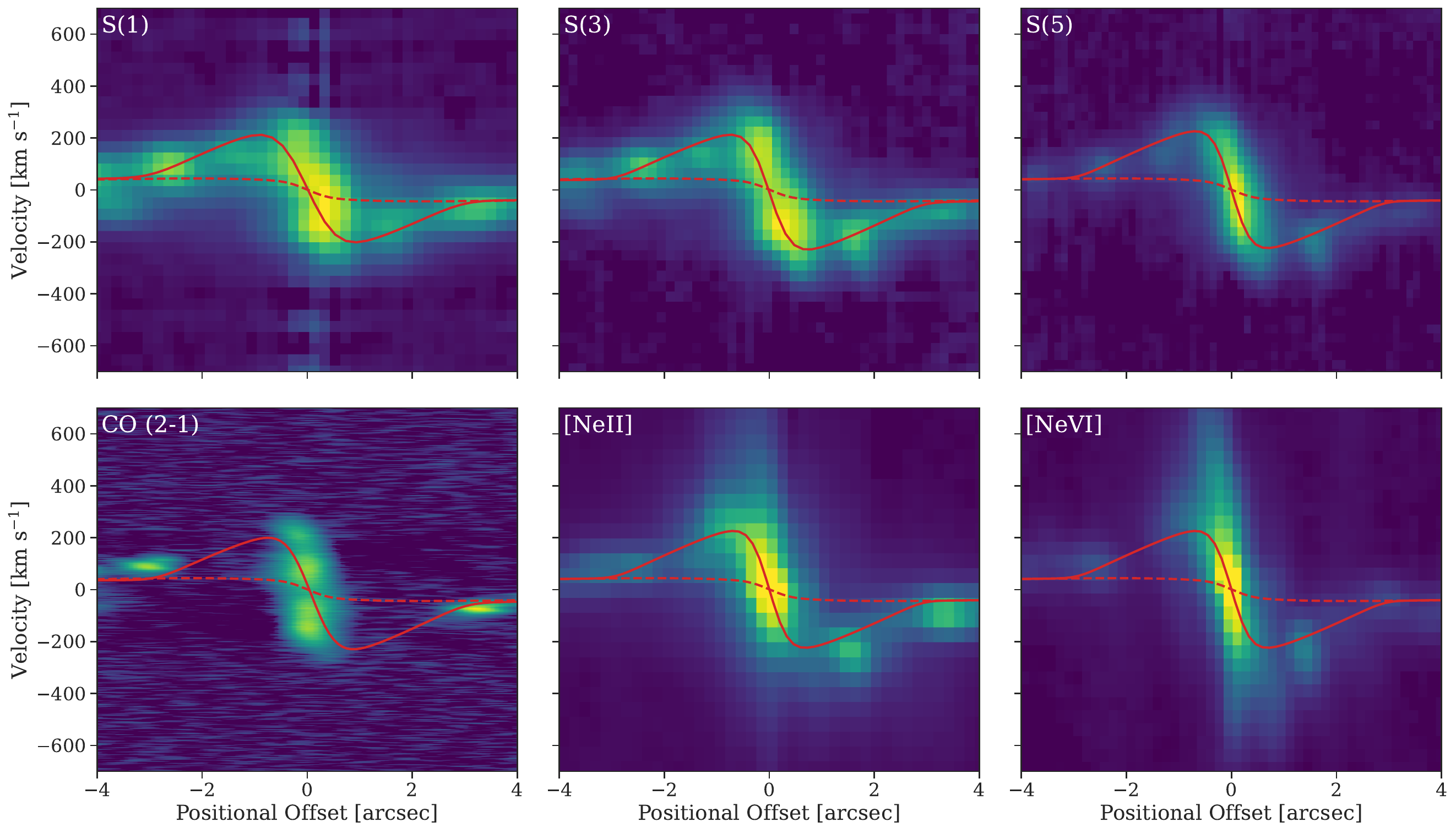}
    \caption{Position-velocity diagrams of the H$_2$ 0-0 S(1), S(3), S(5), CO\,(2-1), [Ne\,II], and [Ne\,VI] emission lines. A pseudo-slit width of 1\arcsec at a position angle of -60 deg East of North was used, from which the continuum has been subtracted (although a residual from wiggles in the continuum can be seen in the panel for the S(1) line). Overplotted are the best fit rotating disk model from \cite[][dashed red line]{shi19} and the best fit rotating disk with a planar radial outflow described in Sec.~\ref{subsec:planaroutflow} for the S(5) line (solid red line). At large radii, all lines trace only the pure rotation while the inner regions need an outflow component. The molecular gas kinematics can be modelled well with a planar radial outflow while the ionised gas clearly has a separate high velocity, compact outflow being lifted off the disk. As comparison numbers for the CO line, outside the central 1--2\arcsec\ the circular velocity is around 300~km~s$^{-1}$ while the dispersion is typically 15~km~s$^{-1}$.}
    \label{fig:h2_pvs}
\end{figure*}

\section{Molecular gas kinematics: a terminated outflow}
\label{sec:h2_kin}

In this Section we return to Fig.~\ref{fig:h2_maps}, the key points of which were briefly summarised in Sec.~\ref{subsec:maps}.
We begin by comparing the flux and kinematics maps of the rotational H$_2$ lines to those of the CO\,(2-1) line, and then compare the molecular outflow to the ionised outflow.
We show that the outflow terminates because it decelerates completely, and we assess what mechanisms may be responsible for that.

\subsection{Planar molecular flow}
\label{subsec:planaroutflow}

Nearly all the features and detailed structure visible in the S(1) map (top left panel of Fig.~\ref{fig:h2_maps}) are matched in the CO\,(2-1) map (bottom left panel). 
The only differences are the detection of molecular gas in the ionisation cones where the CO surface brightness is below the detection limit and, perhaps related to this, enhanced S(1) emission in a location we have identified as the `eastern knot' (region 3 in Fig.~\ref{fig:h2aps}).
Together with the similarities of their velocity maps, as well as of the detailed structure in their dispersion maps, this indicates that the S(1) and CO lines are largely tracing the same component of molecular gas.

The higher rotational lines are qualitatively different, as one might expect from the excitation analysis in Sec.~\ref{sec:h2temp}. 
Most notably, the flux distributions of the S(3) and S(5) lines shown are more prominent in the direction of the ionisation cone and outflow, with the additional hotspot feature (region 2 in Fig.~\ref{fig:h2aps}) becoming apparent to the north-west.

The velocity fields of all the H$_2$ lines are similar to that of the CO line, showing apparent rotation but with a twist in the zero-velocity contour.
A simple disk model, tracing the rotational velocity component, was fitted to the CO map on large scales by \cite{shi19}.
The right-hand column of Fig.~\ref{fig:h2_maps} shows the velocity fields after subtracting this disk model, and clearly reveals a residual in the direction of the ionisation cone indicative of outflowing gas.
We explore this phenomenon in Fig.~\ref{fig:h2_pvs}.
Each panel contains a position-velocity (PV) plot extracted along the direction of the outflow.
Overplotted as a dashed red line is the line-of-sight velocity of the rotation in the disk model.
At the edges of the extracted region, at $\pm$4\arcsec\ corresponding to the radius of the star forming circumnuclear ring, the observed velocities match the model.
But at smaller radii there is a clear discrepancy in the H$_2$ and CO lines, with a projected velocity of up to $\sim$200~km~s$^{-1}$, tracing the molecular outflow.
This can be matched very well if one adds a planar radial component $v_{rad}$ to the disk model (noting that this is the bulk velocity of the clouds, and is different to the shock velocity within the clouds discussed previously).
The solid red line in the panels shows the resulting line-of-sight velocity for 
\begin{equation}
   \label{eqn:vrad}
    v_{rad} = \begin{array}{ll}
    v_{max}~r/r_0 & \mathrm{if~r < r_0} \\ 
    v_{max}~(r_{out}-r)/(r_{out}-r_0) & \mathrm{if~r_0 < r < r_{out}} \\
    \end{array} 
\end{equation}
where the radial velocity $v_{rad}$ reaches its maximum $v_{max} = 400$~km~s$^{-1}$ at $r_0=50$~pc, before decreasing back to $v_{rad}=0$ at $r_{out} = 750$~pc.
The differing central slopes are due to the different beam smearing impacts of the PSF over the wide wavelength range of the lines shown.
That this simple prescription matches the observed PV plots reasonably well suggests that there is a finite extended acceleration zone out to about 50~pc, and beyond that the gas gradually decelerates over a distance of $\sim$700~pc.
Because, the outflow is oriented so close to the disk plane (see Sec.~\ref{sec:circum}), and because the velocity trajectory re-joins that of the disk rotation, we conclude that the molecular outflow is in the plane of the disk.

This outflow is both modest in velocity and limited in spatial extent, terminating within 1~kpc.
It is very different from the ionised outflow that has been discussed extensively in the literature.
The steep trajectory on the PV diagram can be seen clearly for the [Ne\,VI] line, reaching a projected velocity of $>700$~km\,s$^{-1}$ (compared with 200~km\,s$^{-1}$ for the molecular gas) within 1\arcsec. 
And the line emission fades without indicating any tendency to revert back to the disk rotation velocity, suggesting that the ionised gas is escaping the local (few hundred parsec scale) environment.
It is instead more similar in both velocity and extent to the molecular outflows reported in other galaxies.
\citet{dav14} reported molecular outflow velocities in the range 130--190~km~s$^{-1}$ for several galaxies, but these were based on ro-vibrational H$_2$ lines tracing the hottest component of the molecular gas.
\citet{ram22} analysed the CO\,(2-1) line in seven nearby QSOs, find co-planar outflows with slightly higher velocities in the range 200--350~km~s$^{-1}$ extending out to 400-1300~pc.
Detailed modelling of CO\,(3-2) and CO\,(2-1) in several other galaxies continues the emerging trend of planar molecular outflows at velocities up to $\sim$200~km~s$^{-1}$ on scales of several hundred parsecs.
Specifically, for NGC\,3227, \citet{alo19} report evidence for a planar outflow at 180~km~s$^{-1}$ out to 70~pc; for NGC\,7171, \citet{alo23} find radial outflow velocities as high as 80-100~km~s$^{-1}$, decreasing out towards 900~pc; and for NGC\,5506, \citet{esp24} find more modest radial velocity of 50~km~s$^{-1}$ at 50~pc, decreasing to 25~km~s$^{-1}$ and extending out to 600~pc.
The spatial extents of these outflows and that reported here are comparable, as is the indication that the outflow velocity decreases towards larger radii.
That the peak value we find of 400~km~s$^{-1}$ is higher than for some of these other galaxies may be due to two reasons. The simple two component model we use leads to a high peak at small radii that is reduced by beam smearing (and projection effects) in the data, and a more physical model may have a flatter profile. In addition, we have modelled the outflow specifically in one direction, while in some cases an azimuthally symmetric outflow has been adopted. 
It is currently not clear whether this difference reflects a true disparity in the driving mechanism of the outflow or will be reconciled with additional high resolution data, but it seems likely that it should have an impact on the inferred velocities and outflow rates.

\begin{figure}
    \centering
    \includegraphics[width=8cm]{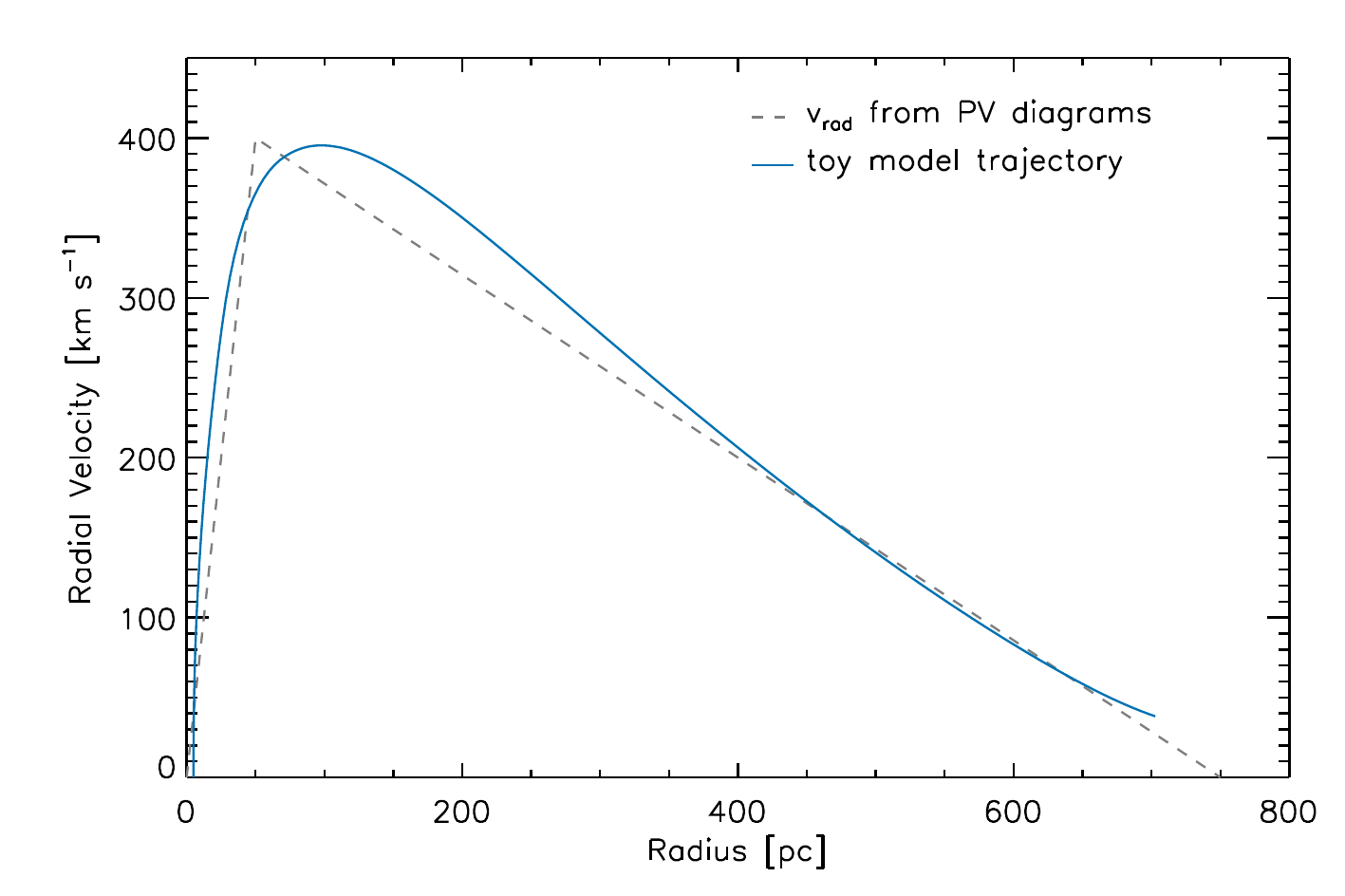}
    \includegraphics[width=7.5cm]{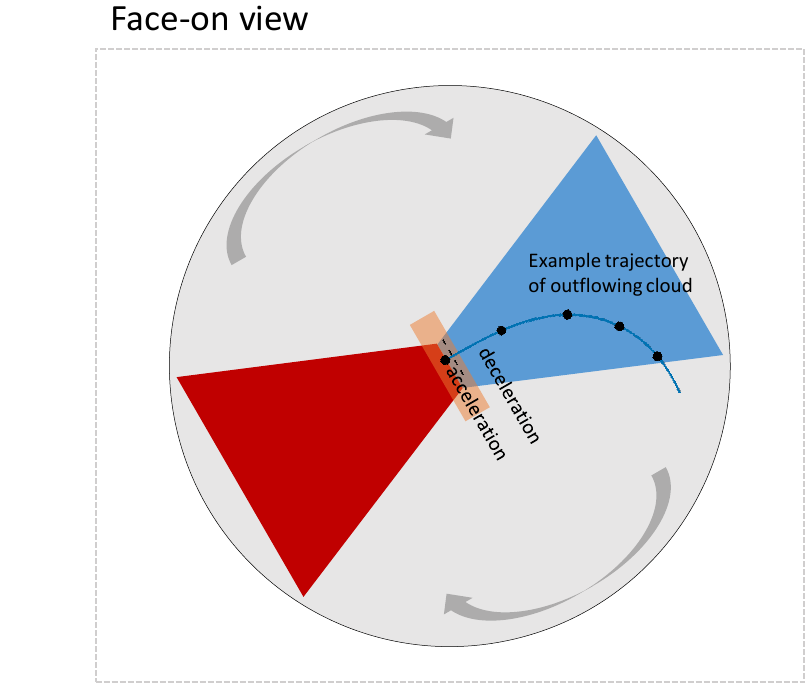}
    \caption{Top panel: radial velocity of the H$_2$ as a function of distance from the centre along the ionisation cone. The dashed grey line represents the simple description given in Eqn.~\ref{eqn:vrad} and Fig.~\ref{fig:h2_pvs}.
    The solid blue line is a toy model described in Sec.~\ref{subsec:decel}, illustrating that this trajectory can be qualitatively assigned to a combination of the increasing gravitational potential and mass-loading.
    Bottom panel: characteristic spatial trajectory of a cloud described by the toy model, where the black circles indicate 0.5~Myr intervals. Initially the motion is radially outwards, but as the cloud decelerates the momentum imparted by mass-loading diverts it to tangential motion.}
    \label{fig:velprofile}
\end{figure}

\subsection{Gas acceleration and deceleration}
\label{subsec:decel}

In this section we look at what might cause the deceleration of the molecular gas.
We focus on two inevitable mechanisms, the increasing gravitational potential and mass-loading due to swept up ambient gas; and consider what happens to a parcel of gas that is accelerated radially and then decelerated while it remains in the outflow region.
The deceleration mechanisms are inter-dependent and so cannot be considered separately.

We can infer the impact of the gravitational potential by deriving the mass profile of the galaxy from the rotation curve of the disk model fitted to the CO velocity field \citep{shi19}.
Throughout most of the disk, gas is supported by rotation. 
But in the outflow, if the gas moves outward, this is no longer the case, and we can calculate the corresponding deceleration term.
At the same time the gas parcel will sweep up ambient gas, which will also cause a deceleration if we assume the collisions are inelastic and that momentum is conserved.
However, the ambient gas has a tangential velocity matching the disk rotation at each radius.
And, again due to momentum conservation, this will tend to increase the tangential velocity of the gas parcel, hence providing greater support against the increased gravitational potential.
It is these interactions that may cause at least some of the shock heating discussed in Sec.~\ref{sec:h2temp}.

In addition to mechanisms that decelerate the gas parcel, there must be an acceleration term causing it to move outwards at all. We adopt a simple prescription that the acceleration is constant out to 25~pc, and then decreases inversely proportionally to the radius. This is not based on any physical principle, but is intended only as a conceptual representation that can match the observed velocity profile.
As noted in Sec.~\ref{sec:circum}, the source of the gas that is being accelerated is within the inner 200~pc scale disk. This means that the outflow is constantly replenished as gas from that inner structure is re-injected back onto circumnuclear scales. This process can continue until the gas reservoir is exhausted and the AGN turns off.

A combination of all three terms is given by the solid blue line in Fig.~\ref{fig:velprofile}. These can provide the rapid acceleration to about 400~km~s$^{-1}$ followed by a roughly linear decline over the subsequent 600~pc.
In this toy model, the gas parcel begins with zero radial and tangential velocity. It is initially accelerated directly outwards, but after about 2~Myr the interaction with the rotating ambient gas means its mass has increased by a factor $\sim2$, transferring angular momentum so that it is increasingly moving tangentially.
This timescale is similar to what one would expect for the crossing time of the ionisation cone, given the rotation velocity at this radius of $\sim$300~km~s$^{-1}$ and the $\sim$600~pc width of the region.
After this time, the gas parcel moves mostly in the tangential direction. The toy model does not explore what happens once the gas leaves the ionisation cone, but indicates that further mass-loading can increase its tangential velocity until it becomes much more similar to the rotation velocity of the ambient disk.

It is not possible to propose a unique solution here, since gas is accelerated over a range of angles and even the highly simplified picture above still has many parameters, without enough constraints.
Instead, our aim is to argue that the observed radial velocity profile can be reproduced by a combination of continuous acceleration in a way that decreases with radius beyond a break point, and which is counteracted by deceleration due to both the increasing gravitational potential and mass-loading by swept up ambient gas;
and that all of these effects are required.

\section{Conclusions}
\label{sec:conc}

As part of a larger study of nearby AGN, we present an analysis of the spatially resolved excitation and kinematics of the rotational H$_2$ lines in the central kiloparsec of the Seyfert~2 galaxy NGC\,5728, observed with JWST/MIRI MRS at 5--28~$\mu$m.
Our main findings are that:

\begin{itemize}

\item H$_2$ is directly detected throughout the ionisation cone, where previous observations had not detected the millimetre CO\,(2-1) line. The gas in the ionisation cone is warmer than in the surrounding regions, or equivalently has a flatter power law index tracing its temperature distribution.

\item Shocks appear to be heating the molecular gas as indicated by the good match between shock models and the H$_2$ level populations, including 0-0, 1-0 and 2-1 vibrational levels. The shock models include UV excitation, which is expected from the AGN. The spatially resolved excitation map supports this, showing that the warmest gas is concentrated (i) along the edges of the ionisation cone into which ambient disk gas is rotating, and (ii) in a region close to where the jet brightens, suggestive of a jet-disk interaction.

\item The molecular gas shows a clear signature of radial flow, which is very different from the outflow traced by the ionised emission lines. It is accelerated over a short distance but then decelerates, suggesting the molecular gas remains in the disk plane. The deceleration mechanisms of this outflow include the effects of the gravitational potential and mass-loading as ambient gas is swept up.

\item A spectro-astrometric signature is measured, indicative of a spatially resolved absorbing structure on scales of $\sim$20~pc. This is slightly misaligned with respect to the 200~pc scale edge-on disk that delineates the direction of the outflow, and may be the reason for asymmetric illumination of the circumnuclear disk at one edge of the ionisation cone.

\end{itemize}

\begin{acknowledgements}

MTL, CP, EKSH, and LZ acknowledge grant support from the Space Telescope Science Institute (ID: JWST-GO-01670.007-A).

AAH, LHM, and MVM research has been funded by grant Nr. PID2021-124665NB-I00 by the Spanish Ministry of Science and Innovation/State Agency of Research MCIN/AEI/ 10.13039/501100011033 and by ``ERDF A way of making Europe''. 

BG-L acknowledges support from the Spanish State Research Agency (AEI-MCINN/10.13039/501100011033) through grants PID2019-107010GB-100 and PID2022-140483NB-C21 PID2022-138560NB-I00, and the Severo Ochoa Program 2020-2023 (CEX2019-000920-S). 

CR acknowledges support from Fondecyt Regular grant 1230345 and ANID BASAL project FB210003.

CRA thanks support by the EU H2020-MSCA-ITN-2019 Project 860744 ``BiD4BESt: Big Data applications for black hole Evolution STudies'' and by project PID2022-141105NB-I00 ``Tracking active galactic nuclei feedback from parsec to kiloparsec scales'', funded by
MICINN-AEI/10.13039/501100011033.

DR acknowledges support from STFC through grants ST/S000488/1 and ST/W000903/1.

EB acknowledges the Mar\'ia Zambrano program of the Spanish Ministerio de Universidades funded by the Next Generation European Union and is also partly supported by grant RTI2018-096188-B-I00 funded by the Spanish Ministry of Science and Innovation/State Agency of Research MCIN/AEI/10.13039/501100011033. 

MPS acknowledges support from grant RYC2021-033094-I funded by MCIN/AEI/10.13039/501100011033 and the European Union NextGenerationEU/PRTR.

MS acknowledges support by the Ministry of Science, Technological Development and Innovation of the Republic of Serbia (MSTDIRS) through contract no. 451-03-66/2024-03/200002 with the Astronomical Observatory (Belgrade).

RD thanks M.~Durr\'e for help with the radio maps and VLA archive. The 6~cm NVAS image of NGC\,5728 was produced as part of the NRAO VLA Archive Survey, which can currently be browsed through http://www.vla.nrao.edu/astro/nvas.

This work is based on observations made with the NASA/ESA/CSA James Webb Space Telescope. The data were obtained from the Mikulski Archive for Space Telescopes at the Space Telescope Science Institute, which is operated by the Association of Universities for Research in Astronomy, Inc., under NASA contract NAS 5-03127 for JWST; and from the European JWST archive (eJWST) operated by the ESAC Science Data Centre (ESDC) of the European Space Agency. These observations are associated with program \#1670.

\end{acknowledgements}

\bibliographystyle{aa}
\bibliography{h2_gas_paper1}

\end{document}